\documentclass[11pt]{article}
\pdfoutput=1

\usepackage[dvipsnames]{xcolor}
\usepackage{graphicx}
\usepackage{subcaption}
\usepackage{epsfig}
\usepackage{epstopdf}
\usepackage[T1]{fontenc}

\DeclareSymbolFont{myletters}{OML}{ztmcm}{m}{it}
\DeclareMathSymbol{\uplambda}{\mathord}{myletters}{"15}

\usepackage{amsfonts}

\usepackage{hyperref}
\hypersetup{
    colorlinks=true,
    linkcolor=blue!70,
    citecolor=RedViolet,
    filecolor=magenta,      
    urlcolor=cyan,
}

\usepackage{color}
\usepackage{float}
\usepackage{multirow}
\usepackage[toc,page]{appendix}

\usepackage{lmodern} 
\usepackage{amsmath}                                                    
\usepackage{amsthm}                                                     
\usepackage{amssymb}
\usepackage{amsfonts}
\usepackage{mathptmx}
\usepackage{slashed}
\usepackage{latexsym}

\numberwithin{equation}{section} 

\newcommand{\newc}{\newcommand}
\newc{\be}{\begin{equation}}
\newc{\ee}{\end{equation}}
\newc{\bg}{\begin{gathered}}
\newc{\eg}{\end{gathered}}
\newc{\tref}[1]{Table \ref{#1}}
\newc{\eref}[1]{Equation \eqref{#1}}
\newc{\su}[1]{$SU(#1)$}
\newc{\bm}[1]{\mathbf{#1}}
\newc{\fref}[1]{Figure \ref{#1}}

\newc{\ra}{\rightarrow}
\newc{\lra}{\leftrightarrow}
\newc{\ov}{\overline}
\newc{\ba}{\begin{eqnarray}}
\newc{\ea}{\end{eqnarray}}
\newc{\mf}{\mathsf}
\newc{\nn}{\nonumber}

\begin{document}

\begin{titlepage}
\thispagestyle{empty}

                \vspace*{0.7cm}

                \begin{center}
    { {\bf \Large{ $SU(5)\times{U(1)'}$ models with a vector-like fermion family }}}
 \\[12mm]
A. Karozas~\footnote{E-mail: \texttt{akarozas@uoi.gr}}, 
G. K. Leontaris~\footnote{E-mail: \texttt{leonta@uoi.gr}} and
I. Tavellaris~\footnote{E-mail: \texttt{i.tavellaris@uoi.gr}}
\end{center}
\vspace*{0.50cm}
\centerline{ \it
Physics Department, University of Ioannina}
\centerline{\it 45110, Ioannina,        Greece}
\vspace*{1cm}
\begin{abstract}

Motivated by experimental measurements indicating deviations from the Standard Model predictions we discuss F-theory inspired models, which, in addition to the three chiral generations contain a vector-like complete fermion family. The analysis takes place in the context of $SU(5)\times U(1)'$ GUT embedded in an $E_8$ covering group which is associated with the (highest) geometric singularity of the elliptic fibration. In this context, the  $U(1)'$  is a linear combination of four abelian factors subjected to appropriate anomaly cancellation conditions. Furthermore, we require universal $U(1)'$ charges for the three chiral families and different ones for the corresponding fields of the vector-like representations. Under the aforementioned assumptions, we find 192 such models which can be classified into five distinct categories with respect to their specific GUT properties. We exhibit representative examples for each such class and construct the superpotential couplings and the fermion mass matrices. We explore the implications of the vector-like states in low energy phenomenology including the predictions regarding the B-meson anomalies. The r\^ole of R-parity violating terms appearing in some particular models of the above construction is also discussed.

\end{abstract}
        \end{titlepage}
        
\setcounter{footnote}{0}
  
\thispagestyle{empty}
\vfill
\newpage
{
  \hypersetup{linkcolor=black}

}

\section{Introduction}

The quest for New Physics (NP) phenomena beyond the Standard Model predictions is a principal and interesting issue. 
Numerous extensions of the Standard Model (SM), including Grand Unified Theories (GUTs) and String Theory derived effective models
incorporate novel ingredients in their spectra. The latter could manifest themselves through exotic interactions and their novel predictions. Amongst the most anticipated  ones are additional neutral gauge bosons, leptoquark states forming couplings with quarks and leptons, 
additional neutral states (such as sterile neutrinos) and 
vector-like families.

Current experimental data  of the Large Hadron Collider (LHC) and 
 elsewhere, on the other hand, provide significant evidence of the existence 
 on possible novel interactions mediated by such exotic states, although
 nothing is conclusive yet. Some well known persisting LHCb data 
 which are in tension with    the SM predictions, for example, are related to various B-meson decay   channels.
In particular measurements of the ratio of the branching ratios $Br(B\to K\mu^+\mu^-)/Br(B\to Ke^+e^-)$
 associated with the
semi-leptonic transitions $b\to s\mu^+\mu^-,b\to se^+e^-$ indicate that  lepton flavor universality is violated \cite{LHCb:2021trn, Aaij:2019wad}. 
Possible explanations of the effect involve leptoquark states,
$Z'$ neutral bosons coupled differently to the three fermion  families
and  vector-like generations \cite{Altmannshofer:2013foa, DAmico:2017mtc, Bifani:2018zmi, Cerri:2018ypt, Kowalska:2019ley, Crivellin:2015mga, Crivellin:2020oup}.

In a previous study \cite{Karozas:2020zvv}, (see also \cite{Romao:2017qnu}) we performed a  systematic analysis of a class of  semi-local   F-theory models with $SU(5)\times U(1)'$ 
gauge symmetry obtained from a  covering  $E_8$  gauge group
through the chain 
\be  
E_8 \supset SU(5)\times SU(5)'\supset SU(5)\times U(1)^4
\supset SU(5)\times U(1)'~,\label{E8to51}
\ee 
where $U(1)'$ stands for any linear combination of
the four abelian factors incorporated in $SU(5)'$.
In this framework, we have derived all possible  solutions of
 the  anomaly-free  $U(1)'$ factors and have shown that many  of these 
 cases  entail  non-universal couplings to the three chiral families.
Next, we considered the case where the spontaneous breaking of the
$U(1)'$ symmetry occurs at a few TeV scale and examined 
the  implications in low energy phenomenology, computing 
observables of several exotic processes in the effective theory.

 Despite the rich structure and the variety of the non-universal 
$U(1)'$ factors, strong lower bounds coming from the $K-\bar K$ system \cite{ParticleDataGroup:2020ssz} on  the mass of the associated 
$Z'$ boson,   far outweigh any observable effects in B-meson
anomalies and the non universal contributions to $Br(B\to K\mu^+\mu^-)/Br(B\to Ke^+e^-)$ are completely
depleted. It was shown that in the so derived
effective F-theory models only the existence of 
additional vector-like families could interpret the
LHCb data \cite{Karozas:2020zvv}.

In the present letter we expand previous work~\cite{Karozas:2020zvv, Romao:2017qnu} 
on F-theory inspired $SU(5)\times U(1)'$ models by including vector-like fermion generations in the low energy spectrum.
More precisely, 
we are interested in models that allow  the existence of a complete family of extra fermions  in addition with the spectrum of the Minimal Supersymmetric Standard Model (MSSM). To avoid severe constraints for the Kaon system, we look for models that the regular MSSM fermion matter fields acquire universal charges under the additional $U(1)'$ symmetry and are chosen to be  different from the corresponding states of the vector-like family. This way the non-universality effects are strictly induced from the considered vector-like states \cite{King:2017anf, Raby:2017igl, Arnan:2019uhr, Freitas:2020ttd}.

The paper is organized as follows: in sections \ref{sec2} and \ref{sec3} we describe the origin of the gauge symmetry of the model, the anomaly cancellation and flux
constraints and define the content in terms of the flux parameters. In section \ref{sec3} in particular 
 all  models with one vector-like  family are sorted out in five classes  distinguished by their $U(1)'$ properties. The phenomenological analysis
 of the models, their superpotential couplings and mass matrices are presented in section \ref{sec4}. Section \ref{sec5} deals with the implications on 
 flavor processes and particularly B-meson anomalies. A short discussion
 is devoted on possible implications of R-parity violating terms in section \ref{sec6}. 
 Summary and conclusions are found in section \ref{sec7}.

\section{Flux constraints for a spectrum with a complete vector-like family}\label{sec2}

 In this section we present 
a short description of the GUT model, focusing mainly  on the basic constraints and characteristics coming from its F-theory embedding. Further technical details  can be found in~\cite{Karozas:2020zvv}.

The (semi-local) F-theory construction in the present work is assumed to originate from an $E_8$ singularity under the  reduction shown in~\eqref{E8to51}. 
The   Cartan generators $Q_k={\rm diag}\{t_1,t_2,t_3,t_4,t_5\},\, k=1,2,3,4$ corresponding to the  four $U(1)$ factors in (\ref{E8to51}), subjected to the $SU(5)$ tracelessness condition $\sum_{i=1}^5t_i=0$,
are taken to be
\begin{eqnarray}
		Q_{a}=\frac{1}{2}
		{\rm diag}(1,-1,0,0,0), 
&&
		Q_{b}=\frac{1}{2\sqrt{3}}{\rm diag}(1,1,-2,0,0),
		\\
		Q_{\psi}=\frac{1}{2\sqrt{6}}{\rm diag}(1,1,1,-3,0),&&
		Q_{\chi}=\frac{1}{2\sqrt{10}}{\rm diag}(1,1,1,1,-4).
\end{eqnarray}
To ensure a tree-level top-quark mass a  $Z_2$ monodromy $t_1\leftrightarrow t_2$ is imposed,  ``breaking''  $U(1)_a$ 
while  leaving invariant  the remaining three abelian  factors. 
In addition, appropriate fluxes \cite{Beasley:2008kw} can be turned on along the remaining 
$U(1)$'s in such a way that some linear combination $U(1)'$ of the abelian factors
remains unbroken at low energies. Thus, the gauge symmetry of  the  effective  model
under consideration is  
\be
G_S=SU(5)\times U(1)^{\prime}\; .\label{F51}
\ee
The  $U(1)^\prime$  factor assumed to be left unbroken in the effective model is a  linear combination of 
the symmetries surviving the monodromy action, namely:
\be
Q^{\prime }= c_1 Q_b + c_2 Q_\psi + c_3 Q_\chi~,\label{Qprime}
\ee
with the coefficients $c_{1}, c_{2}, c_{3} $   satisfying the 
normalization condition
\be
c_1^2+c_2^2+c_3^2 =1~.\label{norm}
\ee
The latter also are subject to anomaly cancellation conditions which have been analysed in detail elsewhere~\cite{Romao:2017qnu,Karozas:2020zvv}.
After imposing the $Z_2$ monodromy, the $10, \overline{10}$
and  $5, \overline{5}$
representations accommodating the massless fields reside on  four 
 matter curves $\Sigma_{10_j}, j=1,2,3,4$ and seven  $\Sigma_{5_i}, i=1,2,\dots, 7 $ respectively~\cite{Dudas:2010zb}. 

 The  $U(1)$ fluxes mentioned above, also 
 determine  the chiralities of  the   $SU(5)$ representations.
 Their effect on the representations of the various matter curves $\Sigma_{10_j}, \Sigma_{5_i} $ 
 can be parametrized   in terms of  integers  $M_j, m_j$
as follows: 
\begin{eqnarray}
n_{10_j}-n_{\overline{10}_j}= m_j\label{M5i}
&&
n_{5_i}-n_{\overline{5}_i}= M_i\label{M10j}~,
\end{eqnarray}
while, in order to accommodate the three fermion families, the chirality condition $
\sum_{j}m_{j}=-\sum_{i}M_{i}=3$ should be imposed. 
Furthermore, turning on a hypercharge flux  $\mathcal{F}_Y$, the $SU(5)_{GUT}$ symmetry is broken  down to $SU(3)\times SU(2)\times U(1)_Y$.
Parametrizing the  hypercharge flux with  integers $N_{i}, N_{j}$ the various multiplicities of the SM representations are given by
\ba
{10}_{t_{j}}=
\left\{\begin{array}{ll}
	n_{{(3,2)}_{\frac 16}}-n_{{(\bar 3,2)}_{-\frac 16}}&=\;m_{j}\\
	n_{{(\bar 3,1)}_{-\frac 23}}-n_{{(
			3,1)}_{\frac 23}}&=\;m_{j}-N_j\\
	n_{(1,1)_{+1}}-n_{(1,1)_{-1}}& =\;m_{j}+N_j\\
\end{array}\right.;\,\;\;\;\;
{5}_{t_{i}}=
\left\{\begin{array}{ll}
	n_{(3,1)_{-\frac 13}}-n_{(\bar{3},1)_{+\frac 13}}&=\;M_{i}\\
	n_{(1,2)_{+\frac 12}}-n_{(1,2)_{-\frac 12}}& =\;M_{i}+N_i\,\; .\\
\end{array}\right.
\label{F5i}
\ea

We start with the flux data and the SM content of each matter curve. For details we refer to our previous work~\cite{Karozas:2020zvv} and here we present only the properties of the complete spectrum as shown in Table \ref{Spectable}. In order to obtain the desired spectrum the following constraints where taken in account.

\begin{table}[tbh!]
 \resizebox{\textwidth}{!}{     
                \begin{tabular}{c|c|c|c|c}
                        \hline\hline
                        Matter Curve  &           $Q^\prime$        &    $N_Y$    &       M     &        SM Content      
                        \\ \hline
                                $\Sigma_{10_{1,\pm t_1}}$    &  $\frac{10 \sqrt{3} c_1+5 \sqrt{6} c_2+3 \sqrt{10} c_3}{60}$   & $-N$ & $m_{1}$  &   $m_{1}Q+(m_{1}+N) u^c +(m_{1}-N)e^c$   
                        \\
                        $\Sigma_{10_{2,\pm t_3}}$   &   $\frac{-20 \sqrt{3} c_1+5 \sqrt{6} c_2+3 \sqrt{10} c_3}{60}$    &    $N_7$    & $m_{2}$  &       $m_{2}Q+(m_{2}- N_7) u^c +(m_{2}+ N_7)e^c$        
                        \\
                        $\Sigma_{10_{3,\pm t_4}}$    &            $\frac{\sqrt{10} c_3-5 \sqrt{6} c_2}{20} $  &    $N_8$    & $m_{3}$  &   $m_{3}Q+(m_{3}- N_8) u^c +(m_{3}+N_8)e^c$       
                        \\
                        $\Sigma_{10_{4,\pm t_5}}$     &     $-\sqrt{\frac{2}{5}} c_3$   &    $N_9$    & $m_{4}$  &   $m_{4}Q+(m_{4}- N_9) u^c +(m_{4}+N_9)e^c$       
                        \\ \hline
                        $\Sigma_{5_{1,(\pm 2t_1)}}$   &     $-\frac{c_1}{\sqrt{3}}-\frac{c_2}{\sqrt{6}}-\frac{c_3}{\sqrt{10}}$     & $N$  & $M_{1}$ & $M_{1}\overline{d^c} + (M_{1}+N)\overline{L}$ 
                        \\
                        $\Sigma_{5_{2,\pm (t_1+t_3)}} $ &    $\frac{5 \sqrt{3} c_1-5 \sqrt{6} c_2-3 \sqrt{10} c_3}{30} $    & $-N$ &  $M_{2}$  &     $M_{2}\overline{d^c} + (M_{2}-N)\overline{L}$    
                        \\
                        $\Sigma_{5_{3,\pm (t_1+t_4)}}$  &    $-\frac{c_1}{2 \sqrt{3}}+\frac{c_2}{\sqrt{6}}-\frac{c_3}{\sqrt{10}}$     & $-N$ &  $M_{3}$  &     $M_{3}\overline{d^c} + (M_{3}- N)\overline{L}$    
                        \\
                        $\Sigma_{5_{4,\pm (t_1+t_5)}}$ & $\frac{-10 \sqrt{3} c_1-5 \sqrt{6} c_2+9 \sqrt{10} c_3}{60} $ & $- N$ &  $M_{4}$  &     $M_{4}\overline{d^c} + (M_{4}-N)\overline{L}$    
                        \\
                        $\Sigma_{5_{5,\pm (t_3+t_4)}} $ &     $\frac{c_1}{\sqrt{3}}+\frac{c_2}{\sqrt{6}}-\frac{c_3}{\sqrt{10}}$      &  $N_7+N_8$  &  $M_{5}$  &     $M_{5}\overline{d^c} + (M_{5}+N_7+N_8)\overline{L}$      
                        \\
                        $\Sigma_{5_{6,\pm (t_3+t_5)}}$ &  $\frac{20 \sqrt{3} c_1-5 \sqrt{6} c_2+9 \sqrt{10} c_3}{60} $  &  $N_7+N_9$  &  $M_{6}$  &     $M_{6}\overline{d^c} + (M_{6}+N_7+N_9)\overline{L}$      
                        \\
                        $\Sigma_{5_{7,\pm (t_4+t_5)}}$ &           $\frac{5 \sqrt{6} c_2+3 \sqrt{10} c_3}{20}$            &  $N_8+N_9$  &  $M_{7}$  &     $M_{7}\overline{d^c} + (M_{7}+N_8+N_9)\overline{L}$      
                        \\ \hline\hline
        \end{tabular}}
\caption{\small{ Matter curves along with their $U(1)'$ charges, flux data and the corresponding SM content. Note that 
		the flux integers satisfy $N=N_{7}+N_{8}+N_{9}$.}}
\label{Spectable}
\end{table}
The spectrum of a  local F-theory model is determined once a set of the above integers -respecting the  aforementioned constraints- is chosen.  Thus, to  start with, we proceed
by  accommodating the Higgs doublet $H_{u}$ on the $\Sigma_{5_1}$ matter curve. Choosing  the associated flux integers to be $M_{1}=0,\,N=1$, it can be observed that  the $H_{u}$ remains in the massless spectrum and at the same time  the down-type color triplet is eliminated. As a consequence of this mechanism~\cite{Beasley:2008kw}, proton decay is sufficiently suppressed. Next, focusing on the $\Sigma_{10_1}$ matter curve, we let $m_1$ vary in $0<m_{1}<3 $. In addition, thanks to the  $Z_{2}$ monodromy~\cite{Dudas:2010zb,King:2010mq}
 already discussed, at least one  diagonal tree-level up-quark Yukawa coupling, $ \lambda_{top} 10_{1}10_{1}5_{1}$, is effectuated in the superpotential $ {\cal W}$. 
Furthermore, in order to ensure exactly one extra family of vector-like fermions, in addition to the condition $\sum_{j}m_{j}=-\sum_{i}M_{i}=3$ 
which fixes the number of chiral families to three, we  also  impose the following conditions into the various flux integers~\cite{Romao:2017qnu,Karozas:2020zvv}:
\be 
\sum_{j=1}^{4}|m_{j}|=
\sum_{i=1}^{7}|M_{i}|=5 ~,
\ee

\be 
|m_{1}+1|+|m_{2}-N_{7}|+|m_{3}-N_{8}|+|m_{4}-N_{9}|=5\; ,
\ee
\be 
|m_{1}-1|+|m_{2}+N_{7}|+|m_{3}+N_{8}|+|m_{4}+N_{9}|=5\; , 
\ee
\be 
1+|M_{2}-1|+|M_{3}-1|+|M_{4}-1|+|M_{5}+N_{7}+N_{8}|+|M_{6}+N_{7}+N_{9}|+|M_{7}+N_{9}+N_{8}|=7 ~.
\ee
Except of $m_{1}, M_{1}$ and $N=N_{7}+N_{8}+N_{9}$ which their allowed ranges and values are  subjected to the aforementioned conditions, the remaining  flux parameters are limited as follows:

\noindent
We restric  the $m_{2,3,4}$ flux integers characterizing the number of $Q,\; \overline{Q}$ states in the spectrum,  in the range $[-1,2]$. Since $\Sigma_{10_1}$ matter curve always hosts at least two $u^{c}$'s (due to conditions $M_{1}=0,\,N=1,\, 0<m_{1}<3 $) we bound the other $u^{c}$ multiplicities ($m_{j}-N_{k}$ with $j=2,3,4$ and $k=7,8,9$) to be in the range $[-1,1]$. Similarly, for the multiplicities of the $e^{c}, \bar{e^c}$ states  we impose  $-1\leq{(m_{j}+N_{k})}\leq{3}$ for $j=2,3,4$ and $k=7,8,9$.  

\noindent
In the same way, for the $d^{c}$'s we set the values of the corresponding multiplicities $M_{i}$'s, ($i=2,3,4,5,6,7$) to vary in the range $[-3,1]$, while for the multiplicities of $\overline{L}$'s (see Table \ref{Spectable}) the relations are set to vary in the range $[-2,1]$. We note here that for the latter, in general we could allow for values in the range $[-3,1]$, but this leads to mixing of the vector-like states with the MSSM ones, something that is against our intention to look for models with vector-like $U(1)'$ charges different than the MSSM ones.

\noindent
Implementing  all the restrictions described above, we receive $1728$ flux solutions with one vector-like family in addition to the  three standard chiral families of quarks and leptons.

\section{Classification of the Models}\label{sec3}

In order to determine the $c_{i}$ coefficients and consequently the $U(1)'$ charges  for each model described by the above set of fluxes, we impose anomaly cancellation conditions. In particular we impose only the mixed MSSM-$U(1)'$ anomalies: $\mathcal{A}_{331}$, $\mathcal{A}_{221}$, $\mathcal{A}_{YY1}$ and $\mathcal{A}_{Y11}$. The pure $U(1)'$ cubic anomaly ($\mathcal{A}_{111}$) and gravitational anomalies ($\mathcal{A}_{G}$) can be fixed later by taking into account  the dynamics of the singlet fields that typically appear in F-theory models.  Furthermore, in the quest for  phenomenologically interesting constructions, we shall  confine our  search in cases where the three MSSM families have universal $U(1)'$ charges and only the charges of the vector-like fields will differ. This way, from the resulting 1728 models only 192 of them appear with  this  property. These 192 models fall into five classes with respect to their $SU(5)\times {U(1)'}$ properties. Each class contains models that carry the same charges under the extra $U(1)'$ and they only differ of how the SM states are distributed among the various matter curves. We present one model for each class in  Tables~\ref{tab:classes_fluxes} and \ref{tab:modelsABCDE}.

\begin{table}[H]
        \resizebox{\textwidth}{!}{%
        \begin{tabular}{c|cccc|ccccccc|ccc|ccc}
                \hline\hline
        Model & $m_1$ & $m_2$ & $m_3$ & $m_4$ & $M_1$                                                                             & $M_2$ & $M_3$ & $M_4$ & $M_5$ & $M_6$ & $M_7$ & $N_7$ & $N_8$ & $N_9$                                                                                                                 & $c_1$ & $c_2$ & $c_3$     \\\hline
\textbf{A} & 1 & 2 &1 & -1 & 0 & -1 & 0 & 0 & -1 & -2 & 1 & 1 & 0 & 0 & 0 & $ \frac{\sqrt{15}}{4}$ & $ -\frac{1}{4}$\\
\textbf{B} & 1 & 2 & -1 & 1 & 0 & 0 & 0 & 0 & -1 & -3 & 1 & 1 & 0 & 0 &0 & $ -\frac{1}{2}\sqrt{\frac{15}{34}}$ & $ -\frac{11}{2\sqrt{34}}$\\
\textbf{C} & 2 & 1 & 1 & -1 & 0 & 0 & 0 & 1 & -3 & -1 & 0 & 0 & 1 & 0 & $\frac{\sqrt{3}}{2}$ & $ -\frac{1}{4}\sqrt{\frac{3}{2}} $& $ \frac{1}{4}\sqrt{\frac{5}{2}}$
\\
\textbf{D} & 2 & -1 & 1 & 1 & 0 & -1 & 0 & 1 & -1 & -2 & 0 & 0 & 0 & 1 & $-\frac{1}{2}\sqrt{\frac{5}{6}}$ & $ -\frac{5}{8}\sqrt{\frac{5}{3}}$ & $ \frac{3}{8}$\\ 
\textbf{E} & 1 & -1 & 2 & 1 & 0 & 0 & 1 & 0 & 0 & -1 & -3 & 0 & 1 & 0  &$2\sqrt{\frac{10}{93}}$ & $ -\sqrt{\frac{5}{93}}$ & $\frac{4}{\sqrt{31}}$\\\hline\hline
        \end{tabular}}
        \caption{ \small{Representative flux solutions along with the corresponding $c_{i}$' s for the five class of models A, B, C, D and E.}}
        \label{tab:classes_fluxes}
        \end{table}

\begin{table}[H]
\resizebox{\textwidth}{!}{%
\begin{tabular}{cc||cc||cc||cc||cc}\hline\hline
\multicolumn{2}{c}{Model \textbf{A}} & \multicolumn{2}{c}{Model \textbf{B}} & \multicolumn{2}{c}{Model \textbf{C}} & \multicolumn{2}{c}{Model \textbf{D}} & \multicolumn{2}{c}{Model \textbf{E}} \\\hline
$\sqrt{10}Q'$   & SM                 & $\sqrt{85}Q'$   & SM                 & $Q'$   & SM                 & $\sqrt{10}Q'$   & SM                 & $\sqrt{310}Q'$   & SM   \\\hline
 1/2    & $Q+2u^c$           & -2      & $Q+2u^c$           & 1/4 & $2Q+3u^c +e^{c}$           & -3/4      & $2Q+3u^c +e^{c}$          & 9/2 & $Q+2u^c$       \\
1/2     & $2Q+u^{c}+3e^{c}$  & -2   &  $2Q+u^{c}+3e^{c}$                 & -1/2   &  $Q+u^{c}+e^{c}$                  & -1/2   & $\overline{Q}+\overline{u^{c}}+\overline{e^{c}} $& 11/2   & $\overline{Q}+\overline{u^{c}}+\overline{e^{c}} $                 \\
-2    &  $Q+u^{c}+e^{c}$         & -1/2      &  $\overline{Q}+\overline{u^{c}}+\overline{e^{c}} $  & 1/4    & $Q+2e^{c}$                  & 7/4    & $Q+u^{c}+e^{c}$                  & 9/2      & $2Q+u^{c}+3e^{c}$   \\
-1/2   &     $\overline{Q}+\overline{u^{c}}+\overline{e^{c}} $             & 11/2    &  $Q+u^{c}+e^{c}$                   & 1/4      & $\overline{Q}+\overline{u^{c}}+\overline{e^{c}} $  & -3/4      &$Q+2e^{c}$   & -8    & $Q+u^{c}+e^{c}$      \\\hline
-1      & $H_u$              & 4      & $H_u$              & -1/2      & $H_u$              & 3/2      & $H_u$              & -9      & $H_u$                 \\
1      & $d^{c}+2L$         & -4   & $L$                & -1/4   & $L$                & -1/4   & $d^{c}+2L$                 & -1   & $L$                 \\
-3/2    & $L$                & -3/2      & $L$         & 1/2    & $L$                & 1    & $L$                & -9      & $\overline{d^{c}}$        \\
1   & $L$                & 7/2    & $L$                & 0      &$\overline{d^{c}}$  & 3/2      & $\overline{d^{c}}$                 & -7/2    & $L$                     \\
-3/2    & $d^c$              & -3/2   & $d^c$              & -1/4      &  $3d^{c}+2L$                   & 9/4      & $d^{c}+L$          & 1   & $\overline{L}$                     \\
1   & $2d^{c}+L$               & 7/2      & $3d^{c}+2L$                   & -3/4   & $d^{c}+L$               & -1/4   & $2d^{c}+L$               & -27/2      & $d^{c}+L$       \\
3/2      & $\overline{d^{c}}+\overline{L}$                  &-6    & $\overline{d^{c}}+\overline{L}$   & 0    & $\overline{L}$              &-1    &$\overline{L}$    &   -7/2            &  $3d^{c}+2L$             \\\hline\hline  
\end{tabular}%
}
\caption{\small{The particle content of models A, B, C, D and E using the data from Table \ref{tab:classes_fluxes}.}}
\label{tab:modelsABCDE}
\end{table}

Table \ref{tab:classes_fluxes} presents the flux data along with the respecting  solution\footnote{Note that a `\emph{mirror}' solution subject to $c_{i}\rightarrow{-c_{i}}$ also exists.} for the coefficients $c_{i}$ of each model.  The corresponding $U(1)'$ charges and the spectrum for each model are  presented in Table \ref{tab:modelsABCDE}. Note that models B and C coincide with the models 5 and 7 respectively,  derived in \cite{Romao:2017qnu}. In addition to the fields presented in Table \ref{tab:modelsABCDE} there are also singlet fields with weights $(t_{i}-t_{j})$ that appear in the present F-theory construction\footnote{For a detailed definition of the singlet spectrum of the theory see \cite{Karozas:2020zvv}.}. In the analysis that follows we will denote these singlet fields as $\theta_{t_{i}-t_{j}}=\theta_{ij}$.

\section{Analysis of the Models}\label{sec4}
In the previous section we generated five classes of models which all share a common characteristic. The $U(1)^{\prime}$ charges of the vector-like  states  differ from  the universal $U(1)^{\prime}$ charges of the SM chiral families. Models with this feature  can explain the observed B-meson anomalies provided there is substantial  mixing of the SM fermions with the vector-like exotics.
At the same time, lepton universality is preserved among the three chiral families and severe bounds coming from the Kaon system and other flavour violating processes are not violated.  In the following  we will analyze the models of Table \ref{tab:modelsABCDE} and derive the mass matrices for each model.

\subsection{Model A}
For this case we have chosen  the following set of fluxes : 
\begin{equation*}
m_{1}=m_{3}=-m_{4}=1\;,\;\; m_{2}=2\;,\;\; M_{1}=M_{3}=M_{4}=0\;,\;\;M_{2}=M_{6}=-2\;,\;\;M_{7}=-M_{5}=1~.
\end{equation*}

 The corresponding $U(1)^{\prime}$ charges for the various representations are :
\begin{align*}
&10_{1}:\frac{1}{2}\;,\;\; 10_{2}:\frac{1}{2}\;,\;\;10_{3}:-2\;,\;\;10_{4}:-\frac{1}{2}~,\\
&5_{1}:-1\;,\;\;5_{2}:-1\;,\;\;5_{3}:\frac{3}{2}\;,\;\;5_{4}:-1\;,\;\;5_{5}:\frac{3}{2}\;,\;\;5_{6}:-1\;,\;\;5_{7}:-\frac{3}{2}~,
\end{align*}
\noindent while the $\ov{10}$, $\ov{5}$ representations come with the opposite $U(1)'$ charge. 
We distribute the fermion generations and Higgs fields into matter curves as follows :
\begin{align*}
10_{1}&\longrightarrow{Q_{3}+u_{2,3}^{c}}\;,\;\; 10_{2}\longrightarrow{Q_{1,2}+u_{1}^{c}+e_{1,2,3}^{c}}\;,\;\;10_{3}\longrightarrow{Q_{4}+u_{4}^{c}+e_{4}^{c}\;,\;\;}\ov{10}_{4}\longrightarrow{\overline{Q_{4}}+\overline{u_{4}^{c}}+\overline{e_{4}^{c}}}\;, \\
5_{1}&\longrightarrow{H_u},\;\; \ov{5}_{2}\longrightarrow{d^{c}_{3}+L_{2,3}},\;\; \ov{5}_{3}\longrightarrow{L_{4}},\;\;\ov{5}_{4}\longrightarrow{H_{d}},\;\;\ov{5}_{5}\longrightarrow{d^{c}_{4}},\;\;\ov{5}_{6}\longrightarrow{d^{c}_{1,2}+L_{1}},\;\;{5}_{7}\longrightarrow{\overline{d^{c}_{4}}+\overline{L_{4}}}~.
\end{align*}
Now we can write down the  superpotential and in particular the
 various terms contributing to the fermion mass matrices. 

\noindent
We start with the up-quark sector.
The dominant contributions to the up-type quark masses descend from the following superpotential terms

\begin{equation}
\begin{split}
W &\supset y_{t}10_{1}10_{1}5_{1}+
\frac{y_{1}}{\Lambda}10_{1}10_{2}5_{1}\theta_{13}+
\frac{y_{2}}{\Lambda^{2}}10_{2}10_{2}5_{1}\theta_{13}^{2}+
\frac{y_{3}}{\Lambda}10_{3}10_{1}5_{1}\theta_{14}+
\frac{y_{4}}{\Lambda^{2}}10_{3}10_{2}5_{1}\theta_{13}\theta_{14}\\
&+
\frac{y_{5}}{\Lambda^{2}}10_{3}10_{3}5_{1}\theta_{14}^{2}+
y_{6}10_{2}\ov{10}_{4}\theta_{53}+y_{7}10_{1}\bar{10_{4}}\theta_{51}+y_{8}10_{3}\ov{10}_{4}\theta_{54}+\frac{y_{9}}{\Lambda}\ov{10}_{4}\ov{10}_{4}\ov{5}_{4}\theta_{51}~,
\end{split}
\label{WtA}
\end{equation}	

\noindent where $y_{i}$'s are coupling constant coefficients and $\Lambda$ is a characteristic  high energy scale of the theory. The operators yield the following mass texture :

\be
M_{u} = \left(\begin{array}{ccccc}
y_{2}\vartheta_{13}^{2}v_{u} & y_{2}\vartheta_{13}^{2}v_{u} &y_{1}\vartheta_{13}v_{u} &y_{4}\vartheta_{13}\vartheta_{14}v_{u}&y_{6}\theta_{43} \\

y_{1}\vartheta_{13}v_{u}&y_{1}\vartheta_{13}v_{u} &\epsilon y_tv_{u} &y_{3}\vartheta_{14}v_{u}&y_{7}\theta_{51}\\

y_{1}\vartheta_{13}v_{u}&y_{1}\vartheta_{13}v_{u}&y_t v_{u}&y_{3}\vartheta_{14}v_{u}&y_{7}\theta_{51}\\

y_{4}\vartheta_{13}\vartheta_{14}v_{u}&y_{4}\vartheta_{13}\vartheta_{14}v_{u}&y_{3}\vartheta_{14}v_{u}&y_{5}\vartheta_{14}^{2}v_{u}&y_{8}\theta_{54}\\

y_{6}\theta_{53}&y_{6}\theta_{53}&y_{7}\theta_{51}&y_{8}\theta_{54}&y_{9}\vartheta_{51}v_{d}
\end{array}\right)~,
\ee 
\noindent where $v_{u}=\langle{H_{u}}\rangle$, $v_{d}=\langle{H_{d}}\rangle$, $\theta_{ij}=\langle{\theta_{ij}}\rangle$, $\vartheta_{ij}=\langle{\theta_{ij}}\rangle/\Lambda$ and $\varepsilon\ll{1}$ is a suppression factor introduced here to capture local effects of Yukawa couplings descending from a common operator \cite{Cecotti:2009zf, Leontaris:2010zd}.

Next we analyse the couplings of the down-quark and charged lepton sectors.
In the vector-like part of the model, up and down quark sectors share some common superpotential operators. These are given in (\ref{WtA}) with couplings $y_{6}$, $y_{7}$ and $y_{8}$. The remaining dominant terms contributing to the down-type quarks are:
\be
\begin{split}
W &\supset \frac{\kappa_{0}}{\Lambda}10_{1}\ov{5}_{2}\ov{5}_{4}\theta_{41}+
\frac{\kappa_{1}}{\Lambda}10_{1}\ov{5}_{6}\ov{5}_{4}\theta_{45}+\frac{\kappa_{2}}{\Lambda}10_{2}\ov{5}_{2}\ov{5}_{4}\theta_{43}+\frac{\kappa_{3}}{\Lambda^{2}}10_{2}\ov{5}_{6}\ov{5}_{4}\theta_{13}\theta_{45}+\frac{\kappa_{4}}{\Lambda^{2}}10_{2}\ov{5}_{6}\ov{5}_{4}\theta_{43}\theta_{15}\\
&+\kappa_{5}10_{3}\ov{5}_{2}\ov{5}_{4}+
\frac{\kappa_{6}}{\Lambda}10_{3}\ov{5}_{6}\ov{5}_{4}\theta_{15}+
\kappa_{7}10_{1}\ov{5}_{5}\ov{5}_{4}+
\frac{\kappa_{8}}{\Lambda}10_{2}\ov{5}_{5}\ov{5}_{4}\theta_{13}+
\frac{\kappa_{9}}{\Lambda}10_{3}\ov{5}_{5}\ov{5}_{4}\theta_{14}\\
&+\frac{\kappa_{10}}{\Lambda}5_{7}\ov{5}_{2}\theta_{41}\theta_{53}+
\frac{\kappa_{11}}{\Lambda}5_{7}\ov{5}_{2}\theta_{51}\theta_{43}+
\kappa_{12}5_{7}\ov{5}_{6}\theta_{43}+
\kappa_{13}5_{7}\ov{5}_{5}\theta_{53}+\frac{\kappa_{14}}{\Lambda}\ov{10}_{4}5_{7}5_{1}\theta_{53} ,
\end{split}~
\label{Wb}
\ee

\noindent  with $\kappa_{i}$ being coupling constant coefficients.

As regards  the charged lepton sector,
we start with the common operators between bottom quark and charged leptons  which are given in (\ref{Wb}) with couplings $\kappa_{2}$, $\kappa_{3}$, $\kappa_{4}$, $\kappa_{5}$ and $\kappa_{6}$. There are also common operators between the up quark and the charged lepton sector which are given in (\ref{WtA}) with couplings $y_{6}$ and $y_{8}$. All the other contributions for the charged lepton mass matrix descend from the operators
\be 
\lambda_{1}10_{2}\ov{5}_{3}\bar{5}_{4}+\frac{\lambda_{2}}{\Lambda}10_{3}\ov{5}_{3}\ov{5}_{4}\theta_{34}+
\lambda_{3}5_{7}\ov{5}_{6}\theta_{43}+\frac{\lambda_{4}}{\Lambda}5_{7}\ov{5}_{2}\theta_{41}\theta_{53}+
\frac{\lambda_{5}}{\Lambda}5_{7}\ov{5}_{2}\theta_{51}\theta_{43}+\lambda_{6}5_{7}\ov{5}_{3}\theta_{51}+
\frac{\lambda_{7}}{\Lambda}\ov{10}_{4}5_{7}5_{1}\theta_{53}~.
\ee

\noindent where $\lambda_{i}$ denotes coupling constant coefficients.

When the various singlet fields $\theta_{ij}$ acquire vacuum expectation values (VEV),  $\langle \theta_{ij}\rangle\ne 0$, they generate hierarchical non-zero entries 
in the mass matrices of quarks and charged leptons. These VEVs, however, 
are subject to phenomenological requirements.
Such an important  constraint comes from the $\mu$-term which
in principle  can be materialized   through the coupling $\theta_{15}5_1\ov 5_4$. Clearly, to avoid decoupling of the Higgs doublets from the light
spectrum, we must require $\langle \theta_{15}\rangle\approx 0$. 
Consequently, the  mass terms involving $\theta_{15}$ of the 
down and charged leptons can be ignored.

We obtain the following down quark mass matrix:
\begin{small}\be
M_{d} = \left(\begin{array}{ccccc}
\kappa_{3}\vartheta_{13}\vartheta_{45}v_{d} &\kappa_{3}\vartheta_{13}\vartheta_{45}v_{d}&\kappa_{1}\vartheta_{45}v_{d}&0&\kappa_{12}\theta_{43} \\

\kappa_{3}\vartheta_{13}\vartheta_{45}v_{d} & \kappa_{3}\vartheta_{13}\vartheta_{45}v_{d}&\kappa_{1}\vartheta_{45}v_{d}&0&\kappa_{12}\theta_{43}\\

\kappa_{2}\vartheta_{43}v_{d}  &\kappa_{2}\vartheta_{43}v_{d}  & \kappa_{0}\vartheta_{41}v_{d}&\kappa_{5}v_{d}&\kappa_{10}\theta_{41}\vartheta_{53}+\kappa_{11}\theta_{51}\vartheta_{43}\\

\kappa_{8}\vartheta_{13}v_{d} &\kappa_{8}\vartheta_{13}v_{d}&\kappa_{7}v_{d}&\kappa_{9}\vartheta_{14}v_{d}&\kappa_{13}\theta_{53}\\

y_{6}\theta_{53}&y_{6}\theta_{53}&y_{7}\theta_{51}&y_{8}\theta_{54}&\kappa_{14}\vartheta_{53}v_{u}
\end{array}\right)\;.
\ee
\label{dwnA}
\end{small}

The mass texture for the charged leptons has the following form :

\be
M_{e} = \left(\begin{array}{ccccc}
\kappa_{3}\vartheta_{13}\vartheta_{45}v_{d} &\kappa_{2}\vartheta_{43}v_{d}&\kappa_{2}\vartheta_{43}v_{d}&\lambda_{1}v_{d}&y_{6}\theta_{53} \\

\kappa_{3}\vartheta_{13}\vartheta_{45}v_{d} &\kappa_{2}\vartheta_{43}v_{d}&\kappa_{2}\vartheta_{43}v_{d}&\lambda_{1}v_{d}&y_{6}\theta_{53}\\

\kappa_{3}\vartheta_{13}\vartheta_{45}v_{d} &\kappa_{2}\vartheta_{43}v_{d} & \kappa_{2}\vartheta_{43}v_{d}&\lambda_{1}v_{d}&y_{6}\theta_{53}\\

0&\kappa_{5}v_{d}&\kappa_{5}v_{d}&\lambda_{2}\vartheta_{34}v_{d}&y_{8}\theta_{54}\\

\lambda_{3}\theta_{43}&\lambda_{4}\theta_{41}\theta_{53}+\lambda_{5}\theta_{51}\theta_{43}&\lambda_{4}
\theta_{41}\theta_{53}+\lambda_{5}\theta_{51}\theta_{43}&\lambda_{6}\theta_{51}&\lambda_{7}\vartheta_{53}v_{u}
\end{array}\right)
\label{lpnA}
\ee

\subsection{Model B}
The second Model is obtained using  the following set of flux parameters :
\begin{equation*}
m_{1}=-m_{3}=m_{4}=1\;,\;\; m_{2}=2\;,\;\; M_{1}=M_{2}=M_{3}=M_{4}=0\;,\;\;M_{7}=-M_{5}=1\;,\;\;M_{6}=-3~.
\end{equation*}
 The corresponding $U(1)^{\prime}$ for the various matter curves are :
\begin{align*}
&10_{1}:-2\;,\;\; 10_{2}:-2\;,\;\;10_{3}:-\frac{1}{2}\;,\;\;10_{4}:\frac{11}{2}~,\\
&5_{1}:4\;,\;\;5_{2}:4\;,\;\;5_{3}:\frac{3}{2}\;,\;\;5_{4}:-\frac{7}{2}\;,\;\;5_{5}:\frac{3}{2}\;,\;\;5_{6}:-\frac{7}{2}\;,\;\;5_{7}:6\; .
\end{align*}
A workable  distribution of the fermion generations and Higgs fields into matter curves is as follows :
\begin{align*}
10_{1}&\longrightarrow{Q_{3}+u_{2,3}^{c}}\;,\;\; 10_{2}\longrightarrow{Q_{1,2}+u_{1}^{c}+e_{1,2,3}^{c}}\;,\;\;\bar{10_{3}}\longrightarrow{\overline{Q_{4}}+\overline{u_{4}^{c}}+\overline{e_{4}^{c}}\;,\;\;}10_{4}\longrightarrow{Q_{4}+u_{4}^{c}+e_{4}^{c}}\;, \\
5_{1}&\longrightarrow{H_u},\;\; \bar{5}_{2}\longrightarrow{H_{d}},\;\; \bar{5}_{3}\longrightarrow{L_{4}},\;\;\bar{5_{4}}\longrightarrow{L_{3}},\;\;\bar{5}_{5}\longrightarrow{d^{c}_{4}},\;\;\bar{5}_{6}\longrightarrow{d^{c}_{1,2,3}+L_{1,2}},\;\;{5}_{7}\longrightarrow{\overline{d^{c}_{4}}+\overline{L_{4}}}~.
\end{align*}
The $\mu$-term here is realized through the coupling $\theta_{13}5_{1}\ov{5_{2}}$, so we require that $\langle \theta_{13}\rangle$ is very small compared to the other singlet VEV's. This restriction obligates us to take high order terms in account for some couplings. We write down the  various terms that construct the fermion mass matrices starting from the up-quark sector.

The dominant contributions to the up-type quarks descend from the following superpotential terms :

\begin{equation}
\begin{split}
W &\supset y_{t}10_{1}10_{1}5_{1}+
\frac{y_{1}}{\Lambda^{2}}10_{1}10_{2}5_{1}\theta_{14}\theta_{43}+
\frac{y_{2}}{\Lambda^{4}}10_{2}10_{2}5_{1}\theta_{14}^{2}\theta_{43}^{2}+
\frac{y_{3}}{\Lambda}10_{1}10_{4}5_{1}\theta_{15}+
\frac{y_{4}}{\Lambda^{2}}10_{2}10_{4}5_{1}\theta_{13}\theta_{15}\\
&+
\frac{y_{5}}{\Lambda^{2}}10_{4}10_{4}5_{1}\theta_{15}^{2}+
y_{6}10_{1}\ov{10}_{3}\theta_{41}+y_{7}10_{2}\ov{10}_{3}\theta_{43}+y_{8}10_{4}\ov{10}_{3}\theta_{45}+\frac{y_{9}}{\Lambda^{2}}\ov{10}_{3}\ov{10}_{3}\ov{5}_{2}\theta_{41}\theta_{43} \; .
\end{split}
\label{WtB}
\end{equation}	

The operators yield the following mass texture :

\be
M_{u} = \left(\begin{array}{ccccc}
y_{2}\vartheta_{14}^{2}\vartheta_{43}^{2}v_{u} & y_{2}\vartheta_{14}^{2}\vartheta_{43}^{2}v_{u} &y_{1}\vartheta_{14}\vartheta_{43}v_{u} &0&y_{7}\theta_{53} \\

y_{1}\vartheta_{14}\vartheta_{43}v_{u}&y_{1}\vartheta_{14}\vartheta_{43}v_{u} &\epsilon y_tv_{u} &y_{3}\vartheta_{15}v_{u}&y_{6}\theta_{41}\\

y_{1}\vartheta_{14}\vartheta_{43}v_{u}&y_{1}\vartheta_{14}\vartheta_{43}v_{u}&y_t v_{u}&y_{3}\vartheta_{15}v_{u}&y_{6}\theta_{41}\\

0&0&y_{3}\vartheta_{15}v_{u}&y_{5}\vartheta_{15}v_{u}&y_{8}\theta_{45}\\

y_{7}\theta_{43}&y_{7}\theta_{43}&y_{6}\theta_{41}&y_{8}\theta_{45}&y_{9}\vartheta_{41}\vartheta_{43}v_{b}
\end{array}\right)
\ee \\

We continue with the bottom sector of the model. There are common operators between top and bottom sector which are given in (\ref{WtB}) with couplings $y_{6}$, $y_{7}$ and $y_{8}$. The remaining dominant terms contributing to the down-type quarks are:
\be
\begin{split}
W &\supset \frac{\kappa_{0}}{\Lambda}10_{1}\ov{5}_{6}\ov{5}_{2}\theta_{43}+
\frac{\kappa_{1}}{\Lambda^{3}}10_{2}\ov{5}_{6}\ov{5}_{2}\theta_{14}\theta_{43}^{2}+\frac{\kappa_{2}}{\Lambda}10_{1}\ov{5}_{5}\ov{5}_{2}\theta_{53}+\frac{\kappa_{3}}{\Lambda^{4}}10_{2}\ov{5}_{5}\ov{5}_{2}\theta_{14}\theta_{43}^{2}\theta_{54}+\frac{\kappa_{4}}{\Lambda^{2}}10_{4}\ov{5}_{5}\ov{5}_{2}\theta_{14}\theta_{34}\\
&+
\frac{\kappa_{5}}{\Lambda^{2}}10_{4}\ov{5}_{6}\ov{5}_{2}\theta_{43}\theta_{15}+
\kappa_{6}5_{7}\ov{5}_{6}\theta_{43}+
\kappa_{7}5_{7}\ov{5}_{5}\theta_{53}+
\frac{\kappa_{8}}{\Lambda}\ov{10}_{3}5_{7}5_{1}\theta_{43} \; .
\end{split}~
\label{WbB}
\ee

From these operators we obtain the following mass matrix describing the down quark sector:
\be
M_{d} = \left(\begin{array}{ccccc}
\kappa_{1}\vartheta_{14}\vartheta_{43}^{2}v_{d} &\kappa_{1}\vartheta_{14}\vartheta_{43}^{2}v_{d}&\epsilon^{2}\kappa_{0}v_{d}&\kappa_{5}\vartheta_{43}\vartheta_{15}&\kappa_{6}\theta_{43} \\

\kappa_{1}\vartheta_{14}\vartheta_{43}^{2}v_{d} &\kappa_{1}\vartheta_{14}\vartheta_{43}^{2}v_{d}&\epsilon\kappa_{0}v_{d}&\kappa_{5}\vartheta_{43}\vartheta_{15}&\kappa_{6}\theta_{43}\\

\kappa_{1}\vartheta_{14}\vartheta_{43}^{2}v_{d} &\kappa_{1}\vartheta_{14}\vartheta_{43}^{2}v_{d}&\kappa_{0}v_{d}&\kappa_{5}\vartheta_{43}\vartheta_{15}&\kappa_{6}\theta_{43}\\

\kappa_{3}\vartheta_{14}\vartheta_{43}^{2}\vartheta_{54}v_{d} &\kappa_{3}\vartheta_{14}\vartheta_{43}^{2}\vartheta_{54}v_{d}&\kappa_{2}\vartheta_{53}v_{d}&\kappa_{4}\vartheta_{14}\vartheta_{34}v_{d}&\kappa_{7}\theta_{53}\\

y_{7}\theta_{43}&y_{7}\theta_{43}&y_{6}\theta_{41}&y_{8}\theta_{45}&\kappa_{8}\vartheta_{43}v_{u} 
\end{array}\right) \; .
\ee

Regarding the charged lepton sector, the common operators between bottom sector and charged leptons are those in (\ref{WbB}) with couplings $\kappa_{1}$, $\kappa_{5}$, $\kappa_{6}$ and $\kappa_{7}$. We also have common operators between the top and charged lepton sector which are given in (\ref{WtB}) with couplings $y_{7}$ and $y_{8}$. Pure charged lepton contributions descend from the  operators
\be  
\frac{\lambda_{1}}{\Lambda}10_{2}\ov{5}_{4}\ov{5}_{2}\theta_{43}+\frac{\lambda_{2}}{\Lambda}10_{4}\ov{5}_{4}\ov{5}_{2}\theta_{45}+\frac{\lambda_{3}}{\Lambda}10_{2}\ov{5}_{3}\ov{5}_{2}\theta_{53}+\lambda_{4}10_{4}\ov{5}_{3}\ov{5}_{2}+\lambda_{5}5_{7}\ov{5}_{4}\theta_{41}+\lambda_{6}5_{7}\ov{5}_{3}\theta_{51}+\frac{\lambda_{7}}{\Lambda}\ov{10}_{3}5_{7}5_{1}\theta_{43}\; .
\ee
\noindent Collectively all the contributions lead to the following mass matrix :

\be
M_{e} = \left(\begin{array}{ccccc}
\kappa_{1}\vartheta_{14}\vartheta_{43}^{2}v_{d}&\kappa_{1}\vartheta_{14}\vartheta_{43}^{2}v_{d}&\lambda_{1}\vartheta_{43}v_{d}&\lambda_{3}\vartheta_{53}v_{d}&y_{7}\theta_{43} \\

\kappa_{1}\vartheta_{14}\vartheta_{43}^{2}v_{d}&\kappa_{1}\vartheta_{14}\vartheta_{43}^{2}v_{d}&\lambda_{1}\vartheta_{43}v_{d}&\lambda_{3}\vartheta_{53}v_{d}&y_{7}\theta_{43} \\

\kappa_{1}\vartheta_{14}\vartheta_{43}^{2}v_{d}&\kappa_{1}\vartheta_{14}\vartheta_{43}^{2}v_{d}&\lambda_{1}\vartheta_{43}v_{d}&\lambda_{3}\vartheta_{53}v_{d}&y_{7}\theta_{43} \\

\kappa_{5}\vartheta_{13}\vartheta_{45}v_{d}+\kappa_{6}\vartheta_{43}\vartheta_{15}v_{d} &\kappa_{5}\vartheta_{13}\vartheta_{45}v_{d}+\kappa_{6}\vartheta_{43}\vartheta_{15}v_{d} &\lambda_{2}\vartheta_{45}v_{d}&\lambda_{4}v_{d}&y_{8}\theta_{45}\\

\kappa_{7}\theta_{43}&\kappa_{7}\theta_{43}&\lambda_{5}\theta_{41}&\lambda_{6}\theta_{51}&\lambda_{7}\vartheta_{43}v_{u}
\end{array}\right)\; .
\ee


\subsection{Model C}
 
Next, a representative model of the class $C$ is analyzed.
The flux integers along with the corresponding $c_{i}$ coefficients are given in Table \ref{tab:classes_fluxes}.  The resulting  $U(1)^{\prime}$ charges for the various matter curves are

\begin{align*}
&10_{1}:\frac{1}{4}\;,\;\; 10_{2}:-\frac{1}{2}\;,\;\;10_{3}:\frac{1}{4}\;,\;\;10_{4}:-\frac{1}{4}~,\\
&5_{1}:-\frac{1}{2}\;,\;\;5_{2}:\frac{1}{4}\;,\;\;5_{3}:-\frac{1}{2}\;,\;\;5_{4}:0\;,\;\;5_{5}:\frac{1}{4}\;,\;\;5_{6}:\frac{3}{4}\;,\;\;5_{7}:0\; .
\end{align*}

\noindent and we assume the following distribution of the various fermion and Higgs fields into matter curves
\begin{align*}
10_{1}&\longrightarrow{Q_{2,3}+u_{1,2,3}^{c}}+e_{3}^{c}\;,\;\; 10_{2}\longrightarrow{Q_{4}+u_{4}^{c}+e_{4}^{c}}\;,\;\;10_{3}\longrightarrow{Q_{1}+e_{1,2}^{c}\;,\;\;}\ov{10}_{4}\longrightarrow{\overline{Q_{4}}+\overline{u_{4}^{c}}+\overline{e_{4}^{c}}}\;, \\
5_{1}&\longrightarrow{H_u},\;\; \ov{5}_{2}\longrightarrow{L_{1}},\;\; \ov{5}_{3}\longrightarrow{H_d},\;\;{5}_{4}\longrightarrow{\overline{d_{4}^{c}}},\;\;\ov{5}_{5}\longrightarrow{d_{1,2,3}^{c}+L_{2,3}},\;\;\ov{5}_{6}\longrightarrow{d^{c}_{4}+L_{4}},\;\;{5}_{7}\longrightarrow{\overline{L_{4}}}~.
\end{align*}
The $\mu$-term here comes through the coupling $\theta_{14}5_{1}\ov{5}_{3}$ and so, we require $\langle \theta_{14}\rangle\approx 0$. Once again we will consider high order terms for some couplings. Next, we write down the  various superpotential terms leading to the the mass matrices for the up, bottom and charged lepton sector.\\

The dominant contributions to the up-type quarks descend from the following superpotential terms :

\begin{equation}
\begin{split}
W &\supset y_{t}10_{1}10_{1}5_{1}+
\frac{y_{1}}{\Lambda^{2}}10_{1}10_{3}5_{1}\theta_{13}\theta_{34}+
\frac{y_{2}}{\Lambda}10_{1}10_{2}5_{1}\theta_{13}+
\frac{y_{3}}{\Lambda^{2}}10_{3}10_{2}5_{1}\theta_{13}\theta_{14}+
\frac{y_{4}}{\Lambda^{2}}10_{2}10_{2}5_{1}\theta_{13}^{2}\\
&+
y_{5}10_{1}\ov{10}_{4}\theta_{51}+
y_{6}10_{2}\ov{10}_{4}\theta_{53}+y_{7}10_{3}\ov{10}_{4}\theta_{54}+\frac{y_{8}}{\Lambda^{2}}\ov{10}_{4}\ov{10}_{4}\ov{5}_{3}\theta_{51}\theta_{54}~,
\end{split}
\label{WtC}
\end{equation}	

\noindent which yield  the following mass texture :

\be
M_{u} = \left(\begin{array}{ccccc}
y_{1}\vartheta_{13}\vartheta_{34}v_{u} &\eta^{3} y_{t}v_{u} &\eta^{2} y_{t}v_{u} &y_{2}\vartheta_{13}v_{u}&y_{5}\theta_{51} \\

y_{1}\vartheta_{13}\vartheta_{34}v_{u} &\eta^{2} y_{t}v_{u} &\eta y_{t}v_{u} &y_{2}\vartheta_{13}v_{u}&y_{5}\theta_{51}\\

y_{1}\vartheta_{13}\vartheta_{34}v_{u} & \eta y_{t}v_{u} &y_{t}v_{u} &y_{2}\vartheta_{13}v_{u}&y_{5}\theta_{51}\\

0&y_{2}\vartheta_{13}v_{u}&y_{2}\vartheta_{13}v_{u}&y_{4}\vartheta_{13}^{2}v_{u}&y_{6}\theta_{53}\\

y_{7}\theta_{54}&y_{5}\theta_{51}&y_{5}\theta_{51}&y_{6}\theta_{53}&y_{8}\vartheta_{51}\vartheta_{54}v_{b}
\end{array}\right)~,
\ee \\

\noindent where $\eta$ is a small constant parameter describing local Yukawa effects.

There are common operators between top and bottom sector. These are given in (\ref{WtC}) with couplings $y_{5}$, $y_{6}$ and $y_{7}$. The remaining operators contributing to the down-type quarks are:
\be
\begin{split}
W &\supset \frac{k}{\Lambda}10_{1}\ov{5}_{5}\ov{5}_{3}\theta_{54}+
\frac{k_{0}}{\Lambda^{3}}10_{3}\ov{5}_{5}\ov{5}_{3}\theta_{13}\theta_{34}\theta_{54}+k_{1}10_{1}\ov{5}_{6}\ov{5}_{3}+\frac{k_{2}}{\Lambda}10_{3}\ov{5}_{6}\bar{5}_{3}\theta_{14}+\frac{k_{3}}{\Lambda}5_{4}\ov{5}_{5}\theta_{14}\theta_{53}\\
&+\frac{k_{4}}{\Lambda^{2}}10_{2}\ov{5}_{5}\ov{5}_{3}\theta_{14}\theta_{53}+
\frac{k_{7}}{\Lambda^{2}}\ov{10}_{4}5_{4}5_{1}\theta_{14}\theta_{53}+\frac{
k_{9}}{\Lambda}10_{2}\ov{5}_{6}\ov{5}_{3}\theta_{13}+
k_{10}5_{4}\ov{5}_{6}\theta_{13}.
\end{split}\; .
\label{WbC}
\ee

Combining all the terms we obtain the following mass matrix describing the down quark sector:
\be \label{mdc}
M_{d}=\left(
\begin{array}{ccccc}
k_0 \vartheta _{13} \vartheta_{34}\vartheta _{54}  v_d &k \varepsilon ^3 \vartheta _{54}  v_d &k \varepsilon ^2 \vartheta _{54}  v_d & 0 &0 \\
 k_{0} \vartheta _{13} \vartheta_{34}\vartheta _{54}  v_d &k \varepsilon ^2 \vartheta _{54} v_d & k\varepsilon  \vartheta _{54}  v_d &0&0 \\
 k_{0}\vartheta _{13} \vartheta_{34}\vartheta _{54} v_d & k \varepsilon  \vartheta_{54} v_d & k\vartheta _{54}  v_d & 0 &0 \\
 0 & k_1 \xi  v_d & k_1 v_d & k_{9}\vartheta _{13}  v_d &  k_{10}\theta_{13} \\
  y_7 \theta_{54} &  y_5 \xi \theta_{51}  &  y_5 \theta_{51} &  y_6 \theta_{53} & k_{7}\vartheta _{14} \vartheta _{53}  v_u \\
\end{array}
\right)~,
\ee

\noindent where $\epsilon$ and $\xi$ are small constant parameters describing local Yukawa effects.

In the charged lepton sector we have some contributions descending from  terms in (\ref{WbC}). These are the operators with couplings $k, k_{0}$, $k_{1}$, $k_{2}$, $k_{4}$ and $k_{9}$. We also have common operators between top and charged lepton sector which are given in (\ref{WtC}) with couplings $y_{5}$, $y_{6}$ and $y_{7}$. All the other leptonic contributions descend from the following operators
\be 
\begin{split}
W \supset 
&\frac{\lambda_{1}}{\Lambda}10_{3}\bar{5}_{2}\bar{5}_{3}\theta_{54}+\frac{\lambda_{2}}{\Lambda}10_{1}\bar{5}_{2}\bar{5}_{3}\theta_{51}+\frac{\lambda_{3}}{\Lambda}10_{2}\bar{5}_{2}\bar{5}_{3}\theta_{53} \\
&+\frac{\lambda_{4}}{\Lambda}5_{7}\bar{5}_{2}\theta_{41}\theta_{53}+\lambda_{5}5_{7}\bar{5}_{5}\theta_{53}+\lambda_{6}5_{7}\bar{5}_{6}\theta_{43}+\frac{\lambda_{7}}{\Lambda}\bar{10}_{4}5_{7}5_{1}\theta_{53}\;.
\end{split}
\ee
\noindent Hence, the mass texture for the charged leptons has the following form :

\be \label{mec}
M_{e} = \left(\begin{array}{ccccc}
\lambda_{1}\vartheta_{54}v_{d}&k_{0}\vartheta_{13}\vartheta_{34}\vartheta_{54}v_{d}&k_{0}\vartheta_{13}\vartheta_{34}\vartheta_{54}v_{d}&0&y_{7}\theta_{54} \\

\lambda_{1}\vartheta_{54}v_{d}&k_{0}\vartheta_{13}\vartheta_{34}\vartheta_{54}v_{d}&k_{0}\vartheta_{13}\vartheta_{34}\vartheta_{54}v_{d}&0&y_{7}\theta_{54}\\

\lambda_{2}\vartheta_{51}v_{d}&k\vartheta_{54}v_{d}&k\vartheta_{54}v_{d}&k_{1}v_{d}&y_{5}\theta_{51}\\

\lambda_{3}\vartheta_{53}v_{d}&0&0&k_{9}\vartheta_{13}v_{d}&y_{6}\theta_{53}\\

\lambda_{4}\vartheta_{41}\theta_{53}&\lambda_{5}\theta_{53}&\lambda_{5}\theta_{53}&\lambda_{6}\theta_{43}&\lambda_{7}\vartheta_{53}v_{u}
\end{array}\right)~.
\ee

\subsection{Model D}
We now pick out   a model belonging to the fourth class. The   $U(1)^{\prime}$ charges for the various matter curves are :
\begin{align*}
&10_{1}:-\frac{3}{4}\;,\;\; 10_{2}:-\frac{1}{2}\;,\;\;10_{3}:\frac{7}{4}\;,\;\;10_{4}:-\frac{3}{4}~,\\
&5_{1}:\frac{3}{2}\;,\;\;5_{2}:\frac{1}{4}\;,\;\;5_{3}:-1\;,\;\;5_{4}:-\frac{3}{2}\;,\;\;5_{5}:-\frac{9}{4}\;,\;\;5_{6}:\frac{1}{4}\;,\;\;5_{7}:1\; .
\end{align*}
A promising distribution of the fermion generations and Higgs fields into the various matter curves is as follows :
\begin{align*}
10_{1}&\longrightarrow{Q_{2,3}+u_{1,2,3}^{c}+e^{c}_{3}}\;,\;\; \bar{10_{2}}\longrightarrow{\overline{Q_{4}}+\overline{u_{4}^{c}}+\overline{e_{4}^{c}}}\;,\;\;10_{3}\longrightarrow{Q_{4}+u_{4}^{c}+e_{4}^{c}\;,\;\;}10_{4}\longrightarrow{Q_{1}+e_{1,2}^{c}}\;, \\
5_{1}&\longrightarrow{H_u},\;\; \bar{5}_{2}\longrightarrow{d^{c}_{1}+L_{1,2}},\;\; \bar{5}_{3}\longrightarrow{H_{d}},\;\;5_{4}\longrightarrow{\overline{d^{c}_{4}}},\;\;\bar{5}_{5}\longrightarrow{d^{c}_{4}+L_{4}},\;\;\bar{5}_{6}\longrightarrow{d^{c}_{2,3}+L_{3}},\;\;{5}_{7}\longrightarrow{\overline{L_{4}}}~.
\end{align*}
In this case the $\mu$-term is realized  through the coupling $\theta_{14}5_{1}\ov{5}_{3}$, and therefore, as in the previous models we require that the singlet VEV is negligibly small, $\langle \theta_{14}\rangle\approx 0$.

Now, we write down the various superpotential terms of the model that lead to the mass matrices for the top and bottom sector.\\

The dominant contributions to the up-type quarks descend from the following superpotential operators :

\begin{equation}
\begin{split}
W &\supset y_{t}10_{1}10_{1}5_{1}+
\frac{y_{1}}{\Lambda}10_{1}10_{4}5_{1}\theta_{15}+
\frac{y_{2}}{\Lambda^{2}}10_{3}10_{1}5_{1}\theta_{13}\theta_{34}+
\frac{y_{3}}{\Lambda^{2}}10_{3}10_{4}5_{1}\theta_{14}\theta_{15}+
\frac{y_{4}}{\Lambda^{4}}10_{3}10_{3}5_{1}\theta_{13}^{2}\theta_{34}^{2}\\
&+
y_{5}10_{1}\ov{10}_{2}\theta_{31}+
y_{6}10_{4}\ov{10}_{2}\theta_{35}+y_{7}10_{3}\ov{10}_{2}\theta_{34}+\frac{y_{8}}{\Lambda^{2}}\ov{10}_{2}\ov{10}_{2}\ov{5}_{3}\theta_{31}\theta_{34} \; .
\end{split}
\label{WtD}
\end{equation}	

These operators yield the following mass texture :

\be
M_{u} = \left(\begin{array}{ccccc}
y_{1}\vartheta_{15}v_{u} &\epsilon^{3}y_{t}v_{u} &\epsilon^{2}y_{t}v_{u} &y_{2}\vartheta_{13}\vartheta_{34}v_{u}&y_{5}\theta_{31} \\

y_{1}\vartheta_{15}v_{u} &\epsilon^{2}y_{t}v_{u} &\epsilon y_{t}v_{u} &y_{2}\vartheta_{13}\vartheta_{34}v_{u}&y_{5}\theta_{31} \\

y_{1}\vartheta_{15}v_{u} &\epsilon y_{t}v_{u} &y_{t}v_{u} &y_{2}\vartheta_{13}\vartheta_{34}v_{u}&y_{5}\theta_{31} \\

0&y_{2}\vartheta_{13}\vartheta_{34}v_{u}&y_{2}\vartheta_{13}\vartheta_{34}v_{u}&y_{4}\vartheta_{13}^{2}\vartheta_{34}^{2}v_{u}&y_{7}\theta_{34}\\

y_{6}\theta_{35}&y_{5}\theta_{31}&y_{5}\theta_{31}&y_{7}\theta_{34}&y_{8}\vartheta_{31}\vartheta_{34}v_{b}
\end{array}\right)\; .
\ee \\

In close analogy with the previous cases,
the operators with couplings $y_{5}$, $y_{6}$ and $y_{7}$ in (\ref{WtD}) contribute also in the bottom sector. The remaining dominant terms contributing to the down-type quarks are:
\be
\begin{split}
W &\supset y_{b}10_{1}\ov{5}_{6}\ov{5}_{3}+
\frac{\kappa_{1}}{\Lambda}10_{1}\ov{5}_{2}\ov{5}_{3}\theta_{51}+\frac{\kappa_{2}}{\Lambda}10_{4}\ov{5}_{6}\ov{5}_{3}\theta_{15}+\kappa_{3}10_{4}\ov{5}_{2}\bar{5}_{3}+\frac{\kappa_{4}}{\Lambda^{2}}10_{3}\ov{5}_{6}\ov{5}_{3}\theta_{13}\theta_{34}\\
&+\frac{\kappa_{5}}{\Lambda}10_{3}\ov{5}_{2}\ov{5}_{3}\theta_{54}+
\frac{\kappa_{6}}{\Lambda}10_{1}\ov{5}_{5}\ov{5}_{3}\theta_{54}+
\frac{\kappa_{7}}{\Lambda}10_{4}\ov{5}_{5}\ov{5}_{3}\theta_{13}+\frac{\kappa_{8}}{\Lambda^{3}}10_{3}\ov{5}_{5}\ov{5}_{3}\theta_{13}\theta_{34}\theta_{54}+
\kappa_{9}5_{4}\ov{5}_{2}\theta_{53}\\
&+\kappa_{10}5_{4}\ov{5}_{6}\theta_{13}+\frac{\kappa_{11}}{\Lambda}5_{4}\ov{5}_{5}\theta_{13}\theta_{54}+
\frac{\kappa_{12}}{\Lambda^{2}}\ov{10}_{2}5_{4}5_{1}\theta_{13}\theta_{34} .
\end{split}~
\label{WbD}
\ee

Collecting all the terms we obtain the following  mass matrix for the down-quark sector of the model:
\begin{small}\be
M_{d} = \left(\begin{array}{ccccc}
\kappa_{3}v_{d} &\kappa_{1}\vartheta_{51}v_{d}&\kappa_{1}\vartheta_{51}v_{d}&\kappa_{5}\vartheta_{54}v_{d}&\kappa_{9}\theta_{53} \\

\kappa_{2}\vartheta_{15}v_{d} &\epsilon^{2}y_{b}v_{d}&\epsilon y_{b}v_{d}&\kappa_{4}\vartheta_{13}\vartheta_{34}v_{d}&\kappa_{10}\theta_{13}\\

\kappa_{2}\vartheta_{15}v_{d} &\epsilon y_{b}v_{d}& y_{b}v_{d}&\kappa_{4}\vartheta_{13}\vartheta_{34}v_{d}&\kappa_{10}\theta_{13}\\

\kappa_{7}\vartheta_{13}v_{d} &\kappa_{6}\vartheta_{54}v_{d}&\kappa_{6}\vartheta_{54}v_{d}&\kappa_{8}\vartheta_{13}\vartheta_{34}\vartheta_{54}v_{d}&\kappa_{11}\theta_{13}\theta_{54}\\

y_{6}\theta_{35}&y_{5}\theta_{31}&y_{5}\theta_{31}&y_{7}\theta_{34}&\kappa_{12}\vartheta_{13}\vartheta_{34}v_{u}
\end{array}\right)\; .
\ee
\end{small}

We turn now to the charged lepton sector of the model. The operators with couplings $\kappa_{1}$, $\kappa_{2}$, $\kappa_{3}$, $\kappa_{4}$, $\kappa_{5}$, $\kappa_{6}$, $\kappa_{7}$ and $\kappa_{8}$ in (\ref{WbD}) contribute also to the charged lepton mass matrix. We also have common operators between top and charged lepton sector which are given in (\ref{WtD}) with couplings $y_{5}$, $y_{6}$ and $y_{7}$. Additional  contributions descend from the operators
\be  
\lambda_{1}5_{7}\ov{5}_{2}\theta_{41}\theta_{53}+\lambda_{2}5_{7}\ov{5}_{2}\theta_{51}\theta_{43}+\lambda_{3}5_{7}\ov{5}_{6}\theta_{43}+\lambda_{4}5_{7}\ov{5}_{5}\theta_{53}+\lambda_{5}\ov{10}_{2}5_{7}5_{1}\; .
\ee
\noindent Combining all the contributions we receive the following mass matrix for the charged leptons of the model

\be
M_{e} = \left(\begin{array}{ccccc}
\kappa_{3}v_{d}&\kappa_{3}v_{d}&\kappa_{2}\vartheta_{15}v_{d}&\kappa_{7}\vartheta_{13}v_{d}&y_{6}\theta_{35} \\

\kappa_{3}v_{d}&\kappa_{3}v_{d}&\kappa_{2}\vartheta_{15}v_{d}&\kappa_{7}\vartheta_{13}v_{d}&y_{6}\theta_{35}  \\

\kappa_{1}\vartheta_{51}v_{d}&\kappa_{1}\vartheta_{51}v_{d}&y_{\tau}v_{d}&\kappa_{6}\vartheta_{54}v_{d}&y_{5}\theta_{31} \\

\kappa_{5}\vartheta_{54}v_{d}&\kappa_{5}\vartheta_{54}v_{d}&\kappa_{4}\vartheta_{14}v_{d}&\kappa_{8}\vartheta_{13}\vartheta_{34}\vartheta_{54}v_{d}&y_{7}\theta_{34}\\

\lambda_{1}\theta_{41}\theta_{53}+\lambda_{2}\theta_{51}\theta_{43}&\lambda_{1}\theta_{41}\theta_{53}+\lambda_{2}\theta_{51}\theta_{43}&\lambda_{3}\theta_{43}&\lambda_{4}\theta_{53}&\lambda_{5}v_{u}
\end{array}\right)\; .
\ee

 \subsection{Model E}
 
For the fifth and final Model we have the following $U(1)^{\prime}$ charges for the various matter curves:
\begin{align*}
&10_{1}:\frac{9}{2}\;,\;\; 10_{2}:\frac{11}{2}\;,\;\;10_{3}:\frac{9}{2}\;,\;\;10_{4}:-8~,\\
&5_{1}:-9\;,\;\;5_{2}:1\;,\;\;5_{3}:9\;,\;\;5_{4}:\frac{7}{2}\;,\;\;5_{5}:-1\;,\;\;5_{6}:\frac{27}{2}\;,\;\;5_{7}:\frac{7}{2}\; .
\end{align*}
In order to receive realistic mass hierarchies we choose the following distribution of the fermion generations and Higgs fields into matter curves:
\begin{align*}
10_{1}&\longrightarrow{Q_{3}+u_{2,3}^{c}}\;,\;\; \bar{10_{2}}\longrightarrow{\overline{Q_{4}}+\overline{u_{4}^{c}}+\overline{e_{4}^{c}}}\;,\;\;10_{3}\longrightarrow{Q_{1,2}+u_{1}^{c}+e_{1,2,3}^{c}\;,\;\;}10_{4}\longrightarrow{Q_{4}+u^{c}_{4}+e_{4}^{c}}\;, \\
5_{1}&\longrightarrow{H_u},\;\; \bar{5}_{2}\longrightarrow{H_{d}},\;\; 5_{3}\longrightarrow{\overline{d^{c}_{4}}},\;\;\bar{5_{4}}\longrightarrow{L_{3}},\;\;5_{5}\longrightarrow{\overline{L_{4}}},\;\;\bar{5}_{6}\longrightarrow{d^{c}_{4}+L_{4}},\;\;\bar{5}_{7}\longrightarrow{d^{c}_{1,2,3}+L_{1,2}}~.
\label{W5tb}
\end{align*}
With this choice the $\mu$-term  is realized through the coupling $\theta_{13}5_{1}\bar{5}_{2}$ which implies that $\langle\theta_{13}\rangle\approx 0$. With this constraint we write down the various operators for the top and bottom quark sector.\\

We start again with dominant contributions to the up-type quarks. These are 
\begin{equation}
\begin{split}
W &\supset y_{t}10_{1}10_{1}5_{1}+
\frac{y_{1}}{\Lambda}10_{1}10_{3}5_{1}\theta_{14}+
\frac{y_{2}}{\Lambda^{2}}10_{3}10_{3}5_{1}\theta_{14}^{2}+
\frac{y_{3}}{\Lambda^{2}}10_{3}10_{4}5_{1}\theta_{14}\theta_{15}+
\frac{y_{4}}{\Lambda}10_{1}10_{4}5_{1}\theta_{15}\\
&+
\frac{y_{5}}{\Lambda^{2}}10_{4}10_{4}5_{1}\theta_{15}^{2}+ y_{6}10_{3}\ov{10}_{2}\theta_{34}+
y_{7}10_{1}\ov{10}_{2}\theta_{31}+y_{8}10_{4}\ov{10}_{2}\theta_{35}+\frac{y_{9}}{\Lambda^{2}}\ov{10}_{2}\ov{10}_{2}\ov{5}_{2}\theta_{31}^{2}
\end{split}
\label{WtE}
\end{equation}	

\noindent and generate the following mass texture :

\be
M_{u} = \left(\begin{array}{ccccc}
y_{2}\vartheta_{14}^{2}v_{u} & y_{2}\vartheta_{14}^{2}v_{u} &y_{1}\vartheta_{14}v_{u} &y_{3}\vartheta_{14}\vartheta_{15}v_{u}&y_{6}\theta_{34} \\

y_{1}\vartheta_{14}v_{u}&y_{1}\vartheta_{14}v_{u} &\epsilon y_{t}v_{u} &y_{4}\vartheta_{15}v_{u}&y_{7}\theta_{31}\\

y_{1}\vartheta_{14}v_{u}&y_{1}\vartheta_{14}v_{u}&y_{t} v_{u}&y_{4}\vartheta_{15}v_{u}&y_{7}\theta_{31}\\

y_{3}\vartheta_{14}\vartheta_{15}v_{u}&y_{4}\vartheta_{15}v_{u}&y_{4}\vartheta_{15}v_{u}&y_{5}\vartheta_{15}^{2}v_{u}&y_{8}\theta_{35}\\

y_{6}\theta_{34}&y_{6}\theta_{34}&y_{7}\theta_{31}&y_{8}\theta_{35}&y_{9}\vartheta_{31}^{2}v_{b}
\end{array}\right)\; .
\ee 

The operators in \eref{WtE} with couplings $y_{6}$, $y_{7}$ and $y_{8}$ contribute also to the bottom sector of the model. In addition we have the following superpotential terms for the bottom sector:
\be
\begin{split}
W &\supset y_{b}10_{1}\ov{5}_{7}\ov{5}_{2}+
\frac{\kappa_{1}}{\Lambda}10_{3}\ov{5}_{7}\ov{5}_{2}\theta_{14}+\frac{\kappa_{2}}{\Lambda}10_{4}\ov{5}_{7}\ov{5}_{2}\theta_{15}+\frac{\kappa_{3}}{\Lambda}10_{3}\ov{5}_{6}\ov{5}_{2}\theta_{13}+\frac{\kappa_{4}}{\Lambda}10_{1}\ov{5}_{6}\ov{5}_{2}\theta_{43}\\
&+\kappa_{5}\ov{5}_{7}{5}_{3}\theta_{15}+
\frac{\kappa_{6}}{\Lambda}\ov{5}_{6}{5}_{3}\theta_{13}\theta_{45}+
\frac{\kappa_{7}}{\Lambda}\ov{10}_{2}{5}_{3}{5}_{1}\theta_{15}+\frac{\kappa_{8}}{\Lambda^{2}}10_{4}\ov{5}_{6}\ov{5}_{2}\theta_{13}\theta_{45}\; .
\end{split}~
\label{WbE}
\ee

Collectively we obtain the following down quark mass matrix:
\be
M_{d} = \left(\begin{array}{ccccc}
\kappa_{1}\vartheta_{14}v_{d} & \kappa_{1}\vartheta_{14}v_{d}&\epsilon^{2}y_{b}v_{d}&\kappa_{2}\vartheta_{15}v_{d}&\kappa_{5}\theta_{15} \\

\kappa_{1}\vartheta_{14}v_{d}&\kappa_{1}\vartheta_{14}v_{d}&\epsilon y_{b}v_{d}&\kappa_{2}\vartheta_{15}v_{d}&\kappa_{5}\theta_{15}\\

\kappa_{1}\vartheta_{14}v_{d}&\kappa_{1}\vartheta_{14}v_{d}& y_{b}v_{d}&\kappa_{2}\vartheta_{15}v_{d}&\kappa_{5}\theta_{15}\\

\kappa_{3}\vartheta_{13}v_{d} &\kappa_{3}\vartheta_{13}v_{d}&\kappa_{4}\vartheta_{43}v_{d}&\kappa_{8}\vartheta_{43}\vartheta_{15}v_{d}&\kappa_{6}\vartheta_{13}\theta_{45}\\

y_{6}\theta_{34}&y_{6}\theta_{34}&y_{7}\theta_{31}&y_{8}\theta_{35}&\kappa_{7}\vartheta_{15}v_{u}
\end{array}\right)
\ee

The bottom sector shares  common operators between with the charged lepton sector of the model. These are given in (\ref{WbE}) with couplings $\kappa_{1}$ ,$\kappa_{2}$ and $\kappa_{8}$. We also have common operators between top and charged lepton sector which are given in (\ref{WtE}) with couplings $y_{6}$ and $y_{8}$. All the other contributions descend from the operators
\be 
\begin{split}
W &\supset y_{\tau}10_{3}\ov{5}_{4}\ov{5}_{2}+\frac{\lambda_{1}}{\Lambda}10_{4}\ov{5}_{4}\ov{5}_{2}\theta_{45}+\frac{\lambda_{2}}{\Lambda}10_{3}\ov{5}_{6}\ov{5}_{2}\theta_{13}+\lambda_{3}5_{5}\ov{5}_{7}\theta_{35}+\frac{\lambda_{4}}{\Lambda}5_{5}\ov{5}_{4}\theta_{45}\theta_{31}\\
&+\lambda_{5}5_{5}\ov{5}_{6}\theta_{45}+ \frac{\lambda_{6}}{\Lambda}\ov{10}_{2}5_{5}5_{1}\theta_{35}\; .
\end{split}
\ee
Combining the various contributions described so far we end up with the following mass matrix for the charged lepton sector of the model :

\be
M_{e} = \left(\begin{array}{ccccc}
\kappa_{1}\vartheta_{14}v_{d}&\kappa_{1}\vartheta_{14}v_{d}&\epsilon^{2}y_{\tau}v_{d}&\lambda_{2}\vartheta_{13} &y_{6}\theta_{34}\\

\kappa_{1}\vartheta_{14}v_{d}&\kappa_{1}\vartheta_{14}v_{d}&\epsilon y_{\tau}v_{d}&\lambda_{2}\vartheta_{13} &y_{6}\theta_{34}\\

\kappa_{1}\vartheta_{14}v_{d}&\kappa_{1}\vartheta_{14}v_{d}&y_{\tau}v_{d}&\lambda_{2}\vartheta_{13} &y_{6}\theta_{34}\\

\kappa_{2}\vartheta_{15}v_{d}&\kappa_{2}\vartheta_{15}v_{d}&\lambda_{1}\vartheta_{45}v_{d}&\kappa_{8}\vartheta_{15}\vartheta_{43}&y_{8}\theta_{35}\\

\lambda_{3}\theta_{35}&\lambda_{3}\theta_{35}&\lambda_{4}\theta_{31}\theta_{45}&\lambda_{5}\theta_{45}&\lambda_{6}\vartheta_{35}v_{u}
\end{array}\right)\; .
\ee

\section{Flavor violation observables}\label{sec5}

Since the $Z^{\prime}$ gauge boson couples differently with the vector-like fields, new flavor violation phenomena might emerge and other  rare processes could  be amplified provided there is sufficient mixing  of the vector-like fields with the SM matter ones \cite{King:2017anf, Raby:2017igl}
. In order to examine whether the present models can account for the observed LHCb-anomalies we need to determine the unitary transformations that diagonalize the mass matrices of the models described in the previous section. 

Due to the complicated form of the various matrices we diagonalize them perturbatively around some small mixing parameter. We perform this procedure for model A while the analysis for the rest of the models is very similar. A detailed phenomenological investigation will follow in a future publication.

\subsection{Some phenomenological predictions of  model A}

To proceed with the analysis and discuss some phenomenological implications,
first we work out the mass matrices and the mixing for quarks and leptons.

{\bf Quarks:\,} We start with the quark sector of  model A and the matrix for the down quarks. In order to simplify the down quark  mass matrix (\ref{dwnA}) we assume that some terms are very small and that approximately vanish. In particular, we consider that $\kappa_{5}=\kappa_{10}=\kappa_{11}=\kappa_{12}=\kappa_{14}=y_{6}=y_{7}\approx{0}$. We further make the following simplifications 
\[ \kappa_{0}\vartheta_{41}v_{d}=\kappa_{1}\vartheta_{45}v_{d}=m\;,\; \kappa_{2}\vartheta_{43}v_{d}=\alpha m\;,\;\; \kappa_{3}\vartheta_{13}\vartheta_{45}v_{d}=\theta m\;,\;\; \kappa_{9}\vartheta_{14}v_{d}=c\mu\;,\;\; \kappa_{8}\theta_{13}v_{d}=b m\;,\;\;\]
\[ \kappa_{13}\theta_{53}\simeq{y_{8}}\theta_{54}=M,\]

 \noindent where the mass parameter $M$  characterizes the mass scale of the extra vector-like states whilst $m\sim {v_{d}}$  is related to the electroweak scale. We have also assumed that the small Yukawa parameters are identical ($\varepsilon\approx{\xi}$). 
 With the above assumptions the matrix receives the simplified form
 
\ba 
M_d=
\left(
\begin{array}{ccccc}
 \theta  m \xi ^4 & \theta  m \xi  & m \xi ^4 & 0 & 0 \\
 \theta  m \xi  & \theta  m & m \xi ^2 & 0 & 0 \\
 \alpha  m \xi ^2 & \alpha  m & m & 0 & 0 \\
 b m & b m \xi  & 0 & c \mu  & M \\
 0 & 0 & 0 & M & 0 \\
\end{array}
\right)~.
\ea 

\noindent In summary, $m,\;M$ and $\mu$ represent mass parameters while $\alpha ,\;\theta,\;c,\;b$ and $\xi$ are dimensionless coefficients.
 Keeping terms up to first order in $\xi$ the mass  matrix $M_{d}M_{d}^{T}$
 can be written as

\be
M_{d}M_{d}^{T}\approx\left(
\begin{array}{ccccc}
 0 & \theta ^2 m^2 \xi  & \alpha  \theta  m^2 \xi  & 0 & 0 \\
 \theta ^2 m^2 \xi  & \theta ^2 m^2 & \alpha  \theta  m^2 & 2 b \theta  m^2 \xi  & 0 \\
 \alpha  \theta  m^2 \xi  & \alpha  \theta  m^2 & \alpha ^2 m^2+m^2 & \alpha  b m^2 \xi  & 0 \\
 0 & 2 b \theta  m^2 \xi  & \alpha  b m^2 \xi  & b^2 m^2+c^2 \mu ^2+M^2 & c \mu  M \\
 0 & 0 & 0 & c \mu  M & M^2 \\
\end{array}
\right)\; .\label{md2A}
\ee
We observe that ~(\ref{md2A}) can be cast in the form $M^{2}_{d}\approx \mathbf{A}+\xi\;\mathbf{B}$ where:

\begin{align} 
\mathbf{A}=\left(
\begin{array}{ccccc}
 0 & 0 & 0 & 0 & 0 \\
 0 & \theta ^2 m^2 & \alpha  \theta  m^2 & 0 & 0 \\
 0 & \alpha  \theta  m^2 & \alpha ^2 m^2+m^2 & 0 & 0 \\
 0 & 0 & 0 & b^2 m^2+c^2 \mu ^2+M^2 & c \mu  M \\
 0 & 0 & 0 & c \mu  M & M^2 \\
\end{array}
\right)\;,\;\; \mathbf{B}=\left(
\begin{array}{ccccc}
 0 & \theta ^2 m^2 & \alpha  \theta  m^2 & 0 & 0 \\
 \theta ^2 m^2 & 0 & 0 & 2 b \theta  m^2 & 0 \\
 \alpha  \theta  m^2 & 0 & 0 & \alpha  b m^2 & 0 \\
 0 & 2 b \theta  m^2 & \alpha  b m^2 & 0 & 0 \\
 0 & 0 & 0 & 0 & 0 \\
\end{array}
\right) \; .
\end{align}

\noindent The local Yukawa parameter $\xi$ couples the electroweak sector with the heavy vector-like part and can be used as a parturbative mixing parameter. The block-diagonal  matrix $\mathbf{A}$, is the leading order part of the matrix and can be diagonalized by a unitary matrix $V_{b_{L}}^{0}$ as $V_{b_{L}}^{0}\mathbf{A}V_{b_{L}}^{0 T}$. For small values of  the parameter $\alpha$ the eigenvalues of this matrix are written as  
\begin{equation}\label{xievΑ}
\begin{split}
&x_{1}=0,\;\; x_{2}\approx{\theta ^2 \left(m^2-\alpha ^2 m^2\right)},\;\; x_{3}\approx{\alpha ^2 \theta ^2 m^2+\alpha ^2 m^2+m^2},\;\; x_{4}\approx M^2 ,\;\; x_{5}\approx b^2 m^2+M^2~.
\end{split}
\end{equation}
\noindent We observe here that the eigenvalues appear with the desired hierarchy. The corresponding unitary matrix which diagonalizes the matrix $\mathbf{A}$ and returns the eigenvalues \eqref{xievΑ}, is

\begin{align} 
V_{b_{L}}^{0}=\left(
\begin{array}{ccccc}
 1 & 0 & 0 & 0 & 0 \\
 0 & \frac{\alpha ^2 \theta ^2}{2}-1 & \alpha  \theta  & 0 & 0 \\
 0 & \alpha  \theta  & 1-\frac{\alpha ^2 \theta ^2}{2} & 0 & 0 \\
 0 & 0 & 0 & -\frac{c \mu  M}{b^2 m^2} & 1 \\
 0 & 0 & 0 & 1 & \frac{c \mu  M}{b^2 m^2} \\
\end{array}
\right)~.
\end{align}

\noindent The columns of this matrix are the unperturbed  eigenvectors $v_{b_{i}}^{0}$ of the initial matrix.



Now we focus on the corrections to the eigenvectors due to the perturbative part $\xi\mathbf{B}$ which are given by the relation :

\begin{equation}
v_{b_{i}}\approx v_{b_{i}}^{0}+\xi\sum_{j\neq{i}}^{5}\frac{(V_{b_{L}}^{0}\mathbf{B}V_{b_{L}}^{0\dagger})_{ji}}{x_{i}-x_{j}}  v_{b_{j}}^{0}~,
\label{aprox}
\end{equation}

\noindent where the second term displays the $\mathcal{O}(\xi)$ corrections to the basic eigenvectors of the leading order matrix $\mathbf{A}$. The corrected diagonalizing matrices schematically receive the form $V_{b_{L}}=V_{b_{L}}^{0}+\xi V_{b_{L}}^{1}$ and similarly for the up quarks and leptons.  This way the mixing parameter $\xi$ enters in the expressions associated with the various flavor violation observables.

Computing the eigenvectors using the formula (\ref{aprox}), the $\mathcal{O}(\xi)$ corrected unitary matrix is :

\begin{align} 
 V_{b_{L}}\approx\left(
\begin{array}{ccccc}
 1 & -\xi  & 0 & 0 & 0 \\
 -\xi  & \frac{\alpha ^2 \theta ^2}{2}-1 & \alpha  \theta  & \frac{2 b \theta  m^2 \xi }{M^2} & -\frac{2 b c \theta  \mu  m^2 \xi }{M^3} \\
 \alpha  \theta  \xi  & \alpha  \theta  & 1-\frac{\alpha ^2 \theta ^2}{2} & -\frac{\alpha  b m^2 \xi }{M^2} & -\frac{\alpha  b c \mu  m^2 \xi }{M^3} \\
 0 & -\frac{2 c \theta  \mu  \xi }{b M} & \frac{\alpha  c \mu  \xi }{b M} & -\frac{c \mu  M}{b^2 m^2} & 1 \\
 0 & \frac{2 b \theta  m^2 \xi }{M^2} & \frac{\alpha  b m^2 \xi }{M^2} & 1 & \frac{c \mu  M}{b^2 m^2} \\
\end{array}
\right)~.
\end{align}

We assume here that the mixing in the top sector is small and that the main mixing descends from the bottom quark sector.

{\bf Charged Leptons:\,} We turn now to the charged lepton mass matrix \eqref{lpnA}. We notice that some parameters from the top and bottom sector contribute also here so the same assumptions for these parameters will be considered here too. Additionally, we assume that $\lambda_{1}=\lambda_{3}=\lambda_{4}=\lambda_{5}=\lambda_{7}\approx{0}$ and we make  the following simplifications
\[ \lambda_{2}\vartheta_{34}v_{d}=c\; \mu\;,\lambda_{5}\vartheta_{51}\vartheta_{43}=q m\;,\; \lambda_{6}\theta_{51}\approx y_{8}\theta_{54}=M\; ~,\]
 \noindent where the mass parameter $M$  characterizes the vector-like scale and $q$, $c$ are dimensionless parameters.\\
 With these approximations the matrix receives the following form :
 \begin{align}
M_{e}\approx\left(
\begin{array}{ccccc}
 \theta  m & \alpha  m \xi ^4 & \alpha  m \xi ^4 & 0 & 0 \\
 \theta  m \xi ^3 & \alpha  m \xi  & \alpha  m \xi ^2 & 0 & 0 \\
 \theta  m \xi  & \alpha  m \xi ^2 & \alpha  m & 0 & 0 \\
 0 & 0 & 0 & c \mu & M \\
 0 & m q & m \xi  q & M & 0 \\
\end{array}
\right)\; .
 \end{align}

We proceed by perturbatively diagonalizing the lepton square mass matrix
$M_{e}M_{e}^{T}$ ($M_{e}^{2}$ for short) using $\xi$ as the expansion parameter. Keeping up to $\mathcal{O}(\xi)$ terms we write the mass square matrix in the form $M^{2}_{e}\approx \mathbf{A}+\xi\;\mathbf{B}$ where:

\begin{align} 
&\mathbf{A}=\left(
\begin{array}{ccccc}
 \theta ^2 m^2 & 0 & 0 & 0 & 0 \\
 0 & 0 & 0 & 0 & 0 \\
 0 & 0 & \alpha ^2 m^2 & 0 & 0 \\
 0 & 0 & 0 & c^2 \mu ^2+M^2 & c \mu  M \\
 0 & 0 & 0 & c \mu  M & m^2 q^2+M^2 \\
\end{array}
\right)\;,\;\; \\
&\mathbf{B}=\left(
\begin{array}{ccccc}
 0 & 0 & \theta ^2 m^2 \xi  & 0 & 0 \\
 0 & 0 & 0 & 0 & \alpha  m^2 \xi  q \\
 \theta ^2 m^2 \xi  & 0 & 0 & 0 & \alpha  m^2 \xi  q \\
 0 & 0 & 0 & 0 & 0 \\
 0 & \alpha  m^2 \xi  q & \alpha  m^2 \xi  q & 0 & 0 \\
\end{array}
\right)\; .
\end{align}

\noindent The eigenvalues of the dominant part are:

\begin{equation}\label{xievC}
x_{1}=0,\;\; x_{2}=\alpha ^2 m^2,\;\; x_{3}=\theta ^2 m^{2},\;\; x_{4}=M^2,\;\; x_{5}=M^2 + m^2 q^2
\end{equation}

\noindent and the unitary matrix $V_{e_{L}}^{0}$, which diagonalizes the dominant matrix $\mathbf{A}$ is:

\begin{align}
	V^0_{e_{L}}=
\left(
\begin{array}{ccccc}
 0 & 1 & 0 & 0 & 0 \\
 0 & 0 & 1 & 0 & 0 \\
 1 & 0 & 0 & 0 & 0 \\
 0 & 0 & 0 & -1 & \frac{c \mu  M}{m^2 q^2} \\
 0 & 0 & 0 & \frac{c \mu  M}{m^2 q^2} & 1 \\
\end{array}
\right)\; .
\end{align}

The $\mathcal{O}(\xi)$ corrections to the eigenvectors due to the perturbative part $\xi\mathbf{B}$ can be found by applying the relation (\ref{aprox}). Then for the final unitary matrix we obtain 
\begin{align}
V_{e_{L}}\approx\left(
\begin{array}{ccccc}
 0 & 1 & 0 & \frac{\alpha  c \mu  m^2 \xi  q}{M^3} & \alpha  \left(-\frac{c^2 \mu ^2 \xi }{m^2 q^3}-\frac{m^2 \xi  q}{M^2}\right) \\
 -\frac{\xi  \left(\alpha ^2+\theta ^2\right)}{\theta ^2} & 0 & 1 & \frac{\alpha  c \mu  m^2 \xi  q}{M^3} & -\frac{\alpha  m^2 \xi  q}{M^2} \\
 1 & 0 & \xi  & 0 & 0 \\
 0 & \frac{\alpha  c \mu  \xi }{M q} & \frac{\alpha  c \mu  \xi }{M q} & -1 & \frac{c \mu  M}{m^2 q^2} \\
 0 & \frac{\alpha  m^2 \xi  q}{M^2} & \frac{\alpha  m^2 \xi  q}{M^2} & \frac{c \mu  M}{m^2 q^2} & 1 \\
\end{array}
\right)
\end{align}

\noindent where in order to simplify the final result we have assumed  the  series expansions for small $\alpha$, $\theta$ and c  keeping only the dominant terms.

\subsection*{B-meson anomalies at LHCb}
In the presence of a fourth generation where the $U(1)'$- charge assignments of its constituents differ from those of the
SM families, many interesting rear flavor processes are expected to be enhanced and a detailed consideration will appear in
a forthcoming publication. Here we  shall focus only on the B-meson anomaly associated with the $b\to s\ell\ell$ decay and in particular the 
ratio  $R_{K^{(*)}}=$ BR$(B\to K^{(*)} \mu\mu)/$BR($K^{(*)}ee$). Due to 
the non-universal coupling of the $Z^{\prime}$ gauge boson with the vector-like fermions,
 the  $C_{9}^{\mu\mu}$ Wilson coefficient which contributes to the flavor violation transition $b\rightarrow{sll}$, is given by:

\be \label{c9wilson}
C_{9}^{\mu\mu}=-\frac{\sqrt{2}}{4G_{F}}\frac{16\pi^{2}}{e^2}\left(\frac{g^{\prime}}{M_{Z^{\prime}}} \right)^{2}\frac{(Q^{\prime}_{d_{L}})_{23}(Q^{\prime}_{e_{L}})_{22}}{V_{tb}V_{ts}^{*}}~,
\ee

\noindent where the matrices $Q^{\prime}_{d_{L}}$ and $Q^{\prime}_{e_{L}}$ defined as \cite{Langacker:2000ju}
 \be \label{mixingchargematrix}
\mathit{Q}^{\prime}_{f_{L}}\equiv V_{f_L}q^{\prime}_{f_{L}}V_{f_L}^{\dagger}~,
\ee 
\noindent with  $q^{\prime}_{f_{L}}$ being $5\times{5}$ diagonal matrices of $U(1)^{\prime}$ charges\footnote{For a discussion on the effects of complex valued contributions to the Wilson coefficients due to large CP-violation effects, see \cite{Carvunis:2021jga}.}.

The elements $(Q^{\prime}_{d_{L}})_{23}$ and $(Q^{\prime}_{e_{L}})_{22}$  participating in to $C_{9}^{\mu\mu}$ coefficient can be obtained from \eqref{mixingchargematrix} using the diagonalization matrices $V_{f_{L}}$ computed above. We have that

\ba
& (Q^{\prime}_{d_{L}})_{23}\approx-\frac{1}{2} \alpha  \theta  \xi ^2\;\;\;\text{and}\;\;\;(Q^{\prime}_{e_{L}})_{22}\approx -1-\xi^{2} .
\ea

Finally, using the set of values $G_{F}\approx{11.66}\; TeV^{-2}$, $e\approx 0.303$, $V_{tb}\approx 0.99$ and $V_{ts}\approx 0.0404$ we estimate that :
\be 
C_{9}^{\mu\mu}\approx -652.5\;\alpha  \theta  \xi ^2 \left(\frac{g'}{M_{Z^{\prime}}}\right)^2+5220\;\alpha  b^2 \theta   \xi ^2 \left(\frac{g'}{M_{Z^{\prime}}}\right)^2 \left(\frac{m}{M}\right)^{4}\label{c9fin}
\ee

\noindent where the mass parameters $m$, $M$ and $M_{Z^{\prime}}$ displayed in $TeV$ units. \\
\noindent Using an indicative set of values $\alpha \to 0.06,\;\theta \to 0.27,\;\xi \to 0.5,\;m\to 0.1,\;b\to 0.1, \;M\to 1.2$ \; in eq.(\ref{c9fin}), we obtain 
\ba 
C_{9}^{\mu\mu}\approx-2.64 \left(\frac{g'}{M_{Z^{\prime}}}\right)^2.
\ea

\noindent According to the most recent global fits \cite{Alguero:2021anc}, an explanation of the current experimental data requires $C_{9}^{\mu\mu}\approx-0.82$, so in this model the  $Z'$ gauge coupling-mass ratio should be of the order   $ \frac{g'}{(M_{Z^{\prime}}/{\rm TeV})}\approx \frac{1}{2 } $  in order the model to explain the observed $R_{K}$ anomalies. This implies a rather small $Z'$ mass \cite{Dwivedi:2019uqd} unless $g'$ is associated with some strong coupling regime. 
Of course the computation 
of $C_{9}^{\mu\mu}$ is very sensitive to the mass and mixing details of the 
representative model chosen in this example and a comprehensive analysis
of whole set of models will determine whether sufficient mixing effects
can predict the various deviations observed in $B$-meson decays, however this analysis is beyond the goal of the present work.

\section{R-parity violation terms}\label{sec6}

A remarkable observation is that particular R-parity violating (RPV) terms such as $\lambda'_{ijk} L_i Q_j d^c_k$  could
explain the anomalies related to the $b\to s \ell\ell$ flavor violating process \cite{Huang:2015vpt, Trifinopoulos:2018rna, Hu:2020yvs, Bardhan:2021adp}.

In this section we look for possible R-parity violating terms  (RPV) in the 
tree-level superpotential (dubbed here $W^{RPV}_{\rm tree}$) for  the models  $A,B,C,D,E$ presented in Table~\ref{tab:modelsABCDE} and briefly discuss their consequences.
We distinguish the RPV terms in those which couple only the MSSM fields and those which  share Yukawa couplings with extra vector-like families.  
If the  former are present in $W^{RPV}_{\rm tree}$, they lead to hard violations of baryon and/or
lepton numbers and must be suppressed. In F-theory constructions, such 
terms can be eliminated either by judicious flux restrictions piercing the various  matter
curves \cite{CrispimRomao:2016tww}, or by additional (discrete) symmetries emanating from the background geometry of the theory~\cite{Hayashi:2009bt, Antoniadis:2012yk, Chen:2010ts, Kimura:2019qxf}. In section 4 of \cite{Antoniadis:2012yk} there are examples how this R-parity can be built.
On the other hand,  provided certain restrictions and conditions
are fulfilled \cite{Huang:2015vpt}, such couplings may contribute to the B-meson anomalies and other interesting effects like the $(g-2)_{\mu}$ anomaly \cite{Altmannshofer:2020axr,Zheng:2021wnu} without exceeding  baryon and/or lepton number violating  bounds. 

Below, for each  one of the five classes of models, among all possible superpotential couplings we single out the RPV terms.  Taking into account that in all the models presented so far the up-Higgs doublet is isolated at the $5_{1}$ matter curve, the possible RPV operators of the $10\cdot \ov{5}\cdot \ov{5}$ form are :
\ba 
 & \textbf{10}_{1}(\ov{\textbf{5}}_{2}\ov{\textbf{5}}_{7}+\ov{\textbf{5}}_{3}\ov{\textbf{5}}_{6}+\ov{\textbf{5}}_{4}\ov{\textbf{5}}_{5}),\;\;\textbf{10}_{2}\ov{\textbf{5}}_{3}\ov{\textbf{5}}_{4},\;\;\textbf{10}_{3}\ov{\textbf{5}}_{2}\ov{\textbf{5}}_{4},\;\;\textbf{10}_{4}\ov{\textbf{5}}_{2}\ov{\textbf{5}}_{3}\; .
\label{rpv}
 \ea
 Next, we discuss each model separately.

\noindent 
{\bf Model A}.
Using Table~\ref{tab:modelsABCDE} and taking into account (\ref{rpv})  we find  that  the only RPV term of the model is:
\ba
& \textbf{10}_{1}\ov{\textbf{5}}_{3}\ov{\textbf{5}}_{6}\longrightarrow L_{4}Q_{3}d_{1,2}^{c}\;.
\ea

\noindent 
We observe that R-parity violation occurs with terms that involve the vector-like family and there are no terms which have only the three quark and lepton families of the MSSM. However, as recently shown in~\cite{Hu:2020yvs, Bardhan:2021adp}, the coupling $L_{4}Q_{3}d_{2}^{c}$ can have significant contributions to the $b\rightarrow{s\mu\mu}$ process through photonic penguin diagrams.

\noindent 
{\bf Model B}. 
Following the same procedure we found that this model contains the following RPV terms:

\ba 
W_{tree}^{RPV}\supset 
 \textbf{10}_{1}\ov{\textbf{5}}_{4}\ov{\textbf{5}}_{5}
+\textbf{10}_{2}\ov{\textbf{5}}_{3}\ov{\textbf{5}}_{4}\longrightarrow L_{3}Q_{3}d_{4}^{c}+L_{3}L_{4}e^{c}_{1,2,3}\; .
\ea

The first operator here does not contribute to $b\rightarrow{sll}$ process due to the absence of the second generation quark in the coupling. On the other hand the term $L_{3}L_{4}e^{c}_{2}$ which descends from the second operator leads to non-negligible contributions to the anomalous magnetic moment of the muon \cite{Altmannshofer:2020axr}. Combining this with non-zero $Z^{\prime}$ contributions may lead to a sufficient explanation of the $(g-2)_{\mu}$ anomaly. 

\noindent 
{\bf Model  C, D and E}. There are no renormalizable RPV terms in these models. So in this case, an explanation of the observed experimental discrepancies is expected from $Z^{\prime}$ interactions and through the mixing of SM fermions with the extra vector-like states.

\section{Conclusions}\label{sec7}
 	In this article  we have expanded our previous work on F-theory motivated models by performing a scan of all the possible $SU(5)\times{U(1)'}$ semi-local constructions predicting a complete family of  vector-like  exotics. 
We use $U(1)'$ hypercharge flux to obtain the symmetry breaking of the non-abelian part and a $Z_2$ monodromy to guarantee a tree-level top Yukawa coupling. Moroever, we have imposed phenomenological restrictions on the various flux parameters in order to obtain exactly three chiral generations and one vector-like complete family of quarks and leptons.  In addition, demanding anomaly cancellation, we have  
found  that there exist 192 models with universal $U(1)'$ charges for the MSSM families and non-universal for the vector-like  states.
These 192 models fall into five distinct classes with respect to their $SU(5)\times{U(1)'}$ properties, classified as $A,B,C,D,E$ in the analysis. We have presented  one illustrative model for each class, exploring the basic properties, computing the superpotential terms, and constructing the fermion mass matrices.
 For  the models derived in the context of class $A$ in particular, we have exemplified  how these types of models can explain the observed $R_{K}$ anomalies through the mixing with the vector-like  states without violating other flavor violation observables, by virtue of the universal nature of the three  SM families.
 We also discussed the presence of RPV couplings in these type of models and their possible contribution in the observed experimental deviations from the SM predictions. It is worth emphasizing   that due to the flux restrictions and the symmetries of the theory only a restricted number of the possible RPV terms appear in the models. This way, with a careful choice of the flux parameters, it is possible to  interpret such deviation effects while avoiding significant contributions to dangerous proton decay effects. A detailed account of  such new physics phenomena is beyond the present work and is left for a future publication.


\end{document}